\documentclass[a4paper,11pt]{article}
\usepackage{pos}
\usepackage[caption=false]{subfig}

\title{Three-hadron dynamics from lattice QCD}

\author*[a]{Fernando Romero-L\'opez}

\affiliation[a]{Albert Einstein Center, Institute for Theoretical Physics, University of Bern, 3012 Bern, Switzerland}

\emailAdd{fernando.romero-lopez@unibe.ch}

\abstract{ Three-hadron spectroscopy is a key frontier in our understanding of the hadron spectrum. In recent years, significant formal and numerical advances have paved the way for studying three-hadron processes directly from lattice QCD, with outstanding applications including the Roper resonance and the doubly charmed tetraquark. This requires theoretical frameworks that relate finite-volume energies to infinite-volume three-particle scattering amplitudes. In this contribution, I discuss recent progress in formulating such frameworks for generic three-hadron systems, and present numerical results for three-meson systems at maximal isospin with physical quark masses, as well as our recent investigation of the three-body dynamics of the doubly charmed tetraquark, $T_{\rm cc}$. 
}

\FullConference{The 11th International Workshop on Chiral Dynamics (CD2024)\\
 26-30 August 2024\\
Ruhr University Bochum, Germany\\}

\newcommand{\cK}[0]{\mathcal K}

\begin{document}
\maketitle

\section{Introduction}

There are many reasons to study three-hadron dynamics. 
First, it is essential to understand how the complexity of the hadron spectrum emerges from Quantum Chromodynamics (QCD). 
This is especially relevant  in light of recent discoveries of exotic hadrons at experiments such as LHCb~\cite{LHCb-FIGURE-2021-001} and BESIII~\cite{Liu:2023hhl}, see Ref.~\cite{Chen:2022asf} for a review. 
A notable example is the recently discovered doubly-charmed tetraquark $T_{\rm cc}(3875)$~\cite{LHCb:2021vvq,LHCb:2021auc}, which has only one decay channel, $T_{\rm cc}\to DD\pi$. 
Another interesting corner of the hadron spectrum are baryon excitations, whose description generally requires the inclusion of three-body effects. 
For instance, the Roper resonance, which decays to both $N\pi$ and $N\pi\pi$, remains poorly understood~\cite{Burkert:2017djo,Mai:2022eur}. 
Similarly, excited hyperons such as $\Lambda(1405)$ and $\Lambda(1520)$ may also require the inclusion of their three-body decay channels, such as $\Lambda\pi\pi$, for a complete theoretical description.

Beyond understanding the hadron spectrum for its own sake, there are numerous applications in nuclear and high-energy physics. For instance, three-nucleon forces are crucial for accurately describing atomic nuclei~\cite{Hammer:2012id}, and three-baryon interactions involving hyperons have been highlighted for their impact on neutron star properties~\cite{Bombaci:2016xzl,Gerstung:2020ktv}. In the context of precision tests of the Standard Model, experimental studies of weak kaon decays into three pions provide some evidence for CP violation~\cite{NA482:2007ucr,NA482:2010gwp}. 

Motivated by these applications, a decade of theoretical efforts has provided the necessary tools to study three-hadron dynamics using lattice QCD~\cite{Detmold:2008gh,Beane:2007qr,Briceno:2012rv,Polejaeva:2012ut,Hansen:2014eka,Hansen:2015zga,Briceno:2017tce,Hammer:2017uqm,Hammer:2017kms,Mai:2017bge,Briceno:2018aml,Briceno:2018mlh,Pang:2019dfe,Jackura:2019bmu,Blanton:2019igq,Briceno:2019muc,
Romero-Lopez:2019qrt,Pang:2020pkl,Blanton:2020gha,Hansen:2020zhy,Blanton:2020jnm,Romero-Lopez:2020rdq,Hansen:2021ofl,Blanton:2020gmf,Muller:2020vtt,Blanton:2021mih,Muller:2021uur,Muller:2022oyw,Pang:2023jri,Bubna:2023oxo,Briceno:2024txg,Xiao:2024dyw,Hansen:2024ffk,Draper:2023xvu,Feng:2024wyg,Jackura:2023qtp,Muller:2020wjo,Draper:2024qeh,Briceno:2024ehy}; see also recent reviews~\cite{Hansen:2019nir,Rusetsky:2019gyk,Mai:2021lwb,Romero-Lopez:2022usb}. The ultimate goal of these approaches is to provide first-principles predictions of three-hadron scattering amplitudes with controlled uncertainties. By identifying poles in these amplitudes, one can predict the masses, decay widths and couplings of resonances. Indeed, there are already many applications~\cite{Beane:2007es,Detmold:2008fn,Detmold:2008yn,Detmold:2011kw,Mai:2018djl,Horz:2019rrn,Blanton:2019vdk,Culver:2019vvu,Mai:2019fba,Fischer:2020jzp,Hansen:2020otl,Alexandru:2020xqf,Brett:2021wyd,Blanton:2021llb,NPLQCD:2020ozd,Baeza-Ballesteros:2022bsn,Draper:2023boj,Abbott:2023coj,Abbott:2024vhj}, including pioneering works on three-body resonances~\cite{Mai:2021nul,Garofalo:2022pux,Yan:2024gwp}. However, the field is relatively new, and some theoretical issues and many potential applications remain to be explored.

Here, I review the methods and some recent results in three hadron spectroscopy. First, I will discuss the background on obtaining three-particle scattering amplitudes from lattice QCD. Then, I will present two applications: the study of three-meson amplitudes at maximal isospin~\cite{Dawid:2025doq,Dawid:2025zxc}, and the three-body dynamics of the doubly charmed tetraquark~\cite{Hansen:2024ffk,Dawid:2024dgy}. I will also provide a brief outlook of the remaining challenges in three-hadron spectroscopy.

\section{Three-hadron scattering amplitudes from lattice QCD}

While lattice QCD calculations are performed in a finite volume and Euclidean spacetime, scattering amplitudes are defined in infinite volume and real time. Therefore, the study of multihadron systems using lattice QCD relies on an indirect method, as originally proposed by L\"uscher for two-particle systems~\cite{Luscher:1986pf,Luscher:1990ux}. The central idea is that finite-volume energies obtained from lattice QCD can be used as constraints on scattering amplitudes. Specifically, quantum field theories in finite volume with periodic boundary conditions only have a discrete spectrum due to quantized momentum modes, as $\boldsymbol{p} = (2\pi/L) \boldsymbol{n}  $, where $L$ is the box size, and $\boldsymbol{n} \in \mathbb{Z}^3$. Stationary energies deviate from the non-interacting case due to multihadron interactions. Intuitively, particles in a large but finite box are mostly separate, but undergo scatterings that shift the energy away from the energy spectrum of a non-interacting theory.

For two-body scattering, the two-particle quantization condition (QC2) relates finite-volume energies to the two-particle  K matrix, $\cK_2$:
\begin{align}
\det_{\ell m} \Big [ 
F_2\big ( E^*_2, \boldsymbol{P}, L \big )^{-1} + \mathcal{K}_{2}(E_2^*) \Big ] = 0\,,
\label{eq:QC2}
\end{align}
where $F_2$ is a known kinematic function that depends on the box-size $L$ and the total momentum of the system $\boldsymbol{P}$ (referred to as the ``L\"uscher zeta function''). In the QC2, the determinant runs over indices that label two-body partial waves, and values of the two-body center-of-mass (CM) energy, $E_n^*$, for which the determinant vanishes correspond to the quantized finite-volume energies. 

The two-body formalism has been applied to many systems and has become an established tool in lattice QCD spectroscopy, see e.g. the recent discussion in Refs.~\cite{Green:2025rel,MorningstarCD}.\footnote{The HalQCD method is another tool to study two-hadron scattering, see e.g.~Ref.~\cite{Aoki:2025abn}.} Remarkably, direct computation of two-body scattering amplitudes directly at the physical point have been conducted~\cite{Fischer:2020jzp,Fischer:2020yvw,Alexandrou:2023elk,Bruno:2023pde,RBC:2023xqv,Boyle:2024hvv,Boyle:2024grr}, and certain calculations are approaching full control of systematic errors, including discretization effects~\cite{Green:2021qol}.

Three-particle tools are based on the same principle: three-body finite-volume energies are determined by three-particle interactions. However, there are some additional difficulties due to the increased complexity of three-particle scattering. First, three-particle amplitudes can be divergent in certain kinematics due to one-particle exchange processes where the exchanged particle goes on shell. Second, three-hadron interactions, both in finite and infinite volume, depend on off-shell two-body interactions via pairwise rescattering. However, a separation between two and three-body effects is not unique. Because of this, the three-hadron formalism needs to introduce an intermediate, cutoff-dependent quantity to parametrize three-body short-range interactions. In the relativistic field-theoretic (RFT) three-particle formalism~\cite{Hansen:2014eka,Hansen:2015zga}, this is denoted as the divergence-free three particle K matrix, $\cK_{\rm df,3}$.\footnote{See Refs.~\cite{Hammer:2017uqm,Hammer:2017kms} for the non-relativistic EFT approach and Ref.~\cite{Mai:2017bge} for the finite-volume unitarity approach.}

Due to this additional complexity, the three-body formalism is structured as a two-step process. The first step employs the three-body quantization condition (QC3) to constrain the three-particle K matrix from lattice QCD energy levels:
\begin{align}
\det_{k \ell m} \Big [ 
F_3\big ( E^*_3, \boldsymbol{P}, L \big )^{-1} + \mathcal{K}_{\rm df,3}(E_3^*) \Big ] = 0\,,
\label{eq:QC3}
\end{align}
where \( F_3 \) is a known matrix incorporating kinematic effects and two-particle interactions via the two-particle K matrix. The determinant is taken over indices describing three on-shell particles, specifically the two-body angular momentum indices of the ``interacting pair'' and the finite-volume momentum of the third ``spectator'' particle.\footnote{If the three particles are not identical, an additional index indicates the choice of spectator.}  The matrices in the QC3 remain finite because we truncate two-body interactions at a maximal partial wave, \( \ell \leq \ell_{\rm max} \), and impose a cutoff on the spectator momentum, \( k \leq k_{\rm max} \). This cutoff dependence explicitly reflects the unphysical nature of \( \mathcal K_{\rm df, 3} \), as noted earlier. The second step involves solving integral equations that relate the two- and three-particle K matrices to the physical scattering amplitude. These integral equations remove the cutoff dependence from the three-body K matrix, yielding a physical amplitude that is both cutoff-independent and consistent with unitarity~\cite{Briceno:2019muc,Jackura:2019bmu}. Solutions to these integral equations in the RFT framework have been presented in Refs.~\cite{Jackura:2020bsk,Dawid:2023jrj,Dawid:2023kxu,Briceno:2024txg}.

In a typical lattice QCD calculation, one has access to a finite number of energy levels to constrain the functional forms of the K matrices. Thus, a practical remark on the finite-volume formalism is that parametrizations of the K matrices must be assumed to fit the finite-volume spectrum in terms of a few parameters. In typical fits, 5 to 10 parameters are constrained through correlated $\chi^2$ minimization. Several approaches to do this optimization have been discussed in Ref.~\cite{Draper:2023boj}, for instance, a combined fit of two- and three-particle energies.  

The two-particle K matrix is often parametrized using simple polynomial or rational functions of the CM momentum $q^2$. For instance, in the case of two pions, the $s$-wave can be described using  
\begin{equation}
    \frac{q}{M_\pi} \cot \delta_0(q)=\frac{M_\pi E_2^*}{E_2^{* 2}-2z^2 M_\pi^2} \sum_{n=0}^{n_{\max }} B_n\left(\frac{q^2}{M_\pi^2}\right)^n,
\end{equation}
where $B_n$ and $z^2$ are fit parameters, the latter corresponding to the position of the Adler zero.  For the three-particle case, additional considerations are required to parametrize $\mathcal K_{\rm df,3}$. One approach that has been employed is a threshold expansion of $\mathcal K_{\rm df,3}$~\cite{Blanton:2019igq}, in which terms are organized in powers of the distance to the three-particle threshold. The number of allowed terms is constrained by particle-exchange as well as C, P, and T symmetries. For instance, for three identical scalar particles, it takes the form  
\begin{equation}
    M^2 \mathcal{K}_{\mathrm{df}, 3}=\mathcal{K}_0+\mathcal{K}_1 \Delta+\mathcal{K}_2 \Delta^2+\mathcal{K}_{\mathrm{A}} \Delta_{\mathrm{A}}+\mathcal{K}_{\mathrm{B}} \Delta_{\mathrm{B}},
\end{equation}
where $9 M_\pi^2 \Delta = P^2 - 9M_\pi^2$, and $\Delta_{\mathrm{A}}$ and $\Delta_{\mathrm{B}}$ are kinematic functions of the four-momenta of the incoming and outgoing particles---see Ref.~\cite{Blanton:2019igq}.

\section{Two and three mesons at maximal isospin}

The three-body formalism is a relatively recent tool, with the first papers appearing only a decade ago~\cite{Polejaeva:2012ut,Briceno:2012rv}. Consequently, many of its features remain unexplored and require careful numerical investigations. The long-term objective is to apply these tools to systems with three-body resonances, and even in the context of three-body electroweak transitions. However, it is essential to begin with a well-controlled setup, e.g.~weakly interacting systems, such as $3\pi^+$. In these systems, no resonances are present, but statistical signals are typically strong and require relatively lower computational cost, making them an excellent testing ground. Moreover, in certain cases, comparisons with effective theories are possible, providing valuable benchmarks.

In our most recent works~\cite{Dawid:2025doq,Dawid:2025zxc}, we focus on such systems, specifically, three-body systems built of pions or kaons: $3\pi^+, 3K^+,\pi^+\pi^+ K^+$, and $ K^+K^+\pi^+$. We had already explored such systems at heavier-than-physical quark masses~\cite{Blanton:2021llb,Draper:2023boj}, ranging from 200-340 MeV. In the latest update, we included an ensemble with physical quark masses---the E250 CLS ensemble~\cite{Mohler:2017wnb}.

\begin{figure}[th!]
     \centering
    \includegraphics[width=0.95\textwidth]{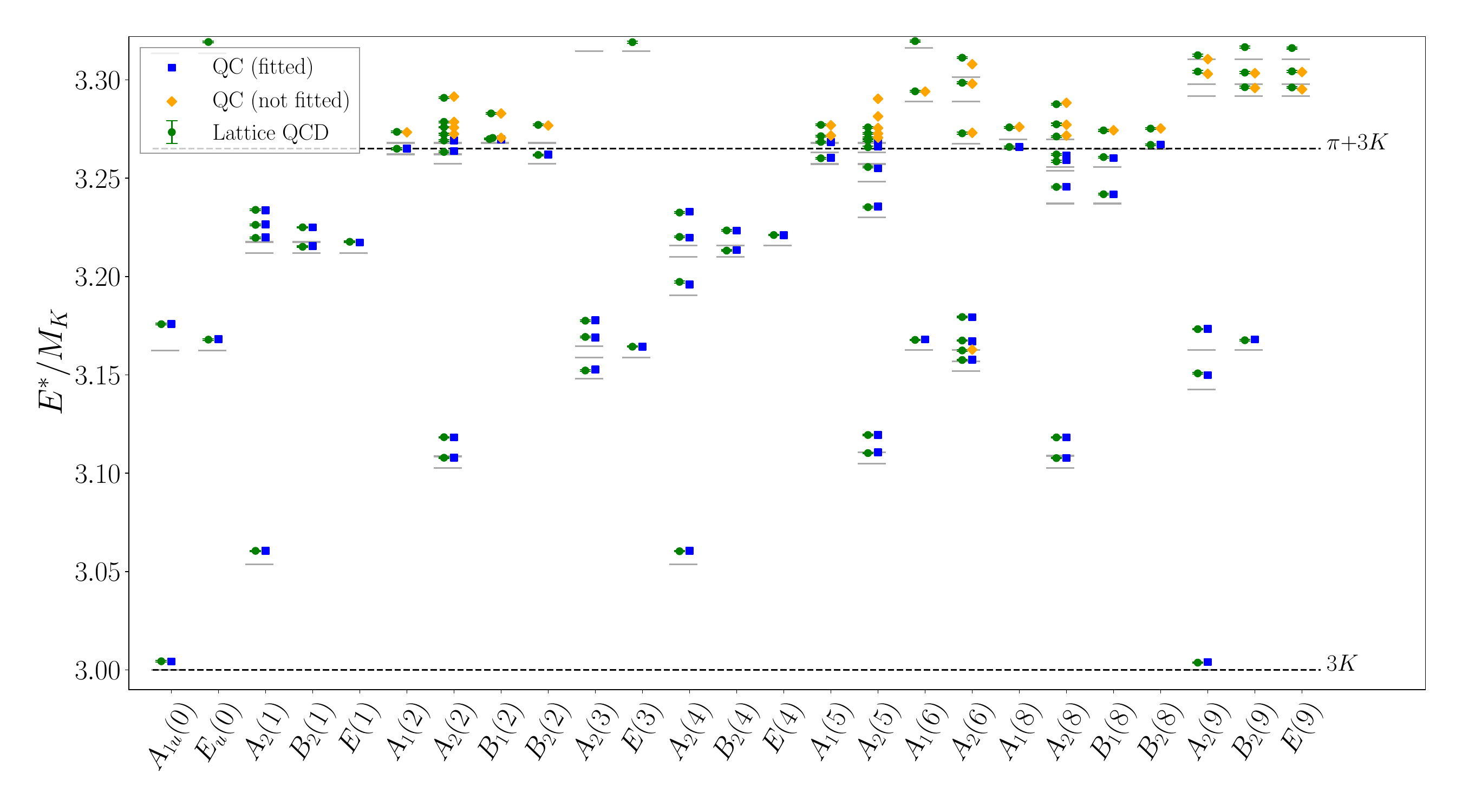}
    \caption{ Finite-volume energies of the $3K^+$ system with physical quark masses in different irreducible representations and finite-volume frames~\cite{Dawid:2025doq,Dawid:2025zxc}.  Green circles are lattice QCD determinations, while blue squares indicate predictions from the three-particle finite-volume formalism for the levels included in the fit, and orange diamonds for levels not included in the fit. The elastic and first inelastic thresholds are shown as black dashed horizontal lines. 
    }
    \label{fig:KKK}
\end{figure}

The calculation begins with the extraction of energies from the lattice QCD Euclidean correlation functions. This is achieved by using a large set of operators with the quantum numbers of the system we aim to investigate. The matrix of correlation functions is analyzed as a generalized eigenvalue problem~\cite{Luscher:1990ck}.\footnote{For alternative novel methods of energy extraction, see Refs.~\cite{Hackett:2024xnx,Wagman:2024rid,Ostmeyer:2024qgu,Hackett:2024nbe,Abbott:2025yhm}.} From the exponential decay of the generalized eigenvalues in Euclidean-time, the low-lying energies of the system can be determined. An example of the outcome of such calculation is provided in Fig.~\ref{fig:KKK} for a system of three kaons with physical quark masses. 

We simultaneously fit the two- and three-body finite-volume energy levels. In the case of Fig.~\ref{fig:KKK}, this means that both $2K^+$ and $3K^+$ energy levels are analyzed together. The extracted parameters are then used to generate predictions from the quantization condition, also displayed in Fig.~\ref{fig:KKK}. Although the formalism is strictly valid only below the first inelastic threshold, its breakdown is not immediate, and the first few energy levels above this inelastic threshold are still well-reproduced.

\begin{figure}[th!]
     \centering
    \includegraphics[width=\textwidth]{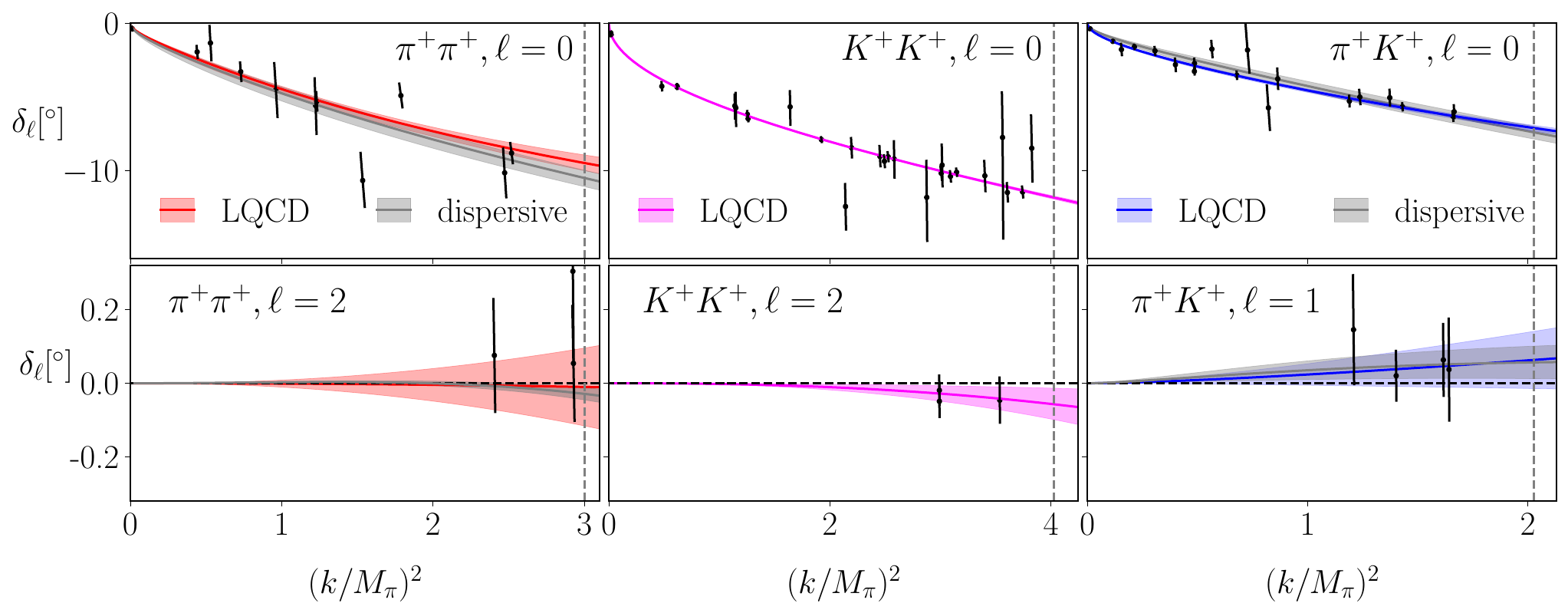}
    \caption{
    Phase shift as a function of the squared scattering momentum in the lowest two partial waves for maximal-isospin two-meson systems with physical quark masses~\cite{Dawid:2025doq,Dawid:2025zxc}. Dispersive results~\cite{Garcia-Martin:2011iqs,Pelaez:2020gnd} are also shown. The vertical grey dashed lines mark the lowest inelastic threshold. Black points are mapped from two-meson energy levels under the assumption that only the lowest partial wave contributes.
    } \label{fig:twobody}
\end{figure}

Given that two- and three-meson energies have been analyzed together, a byproduct is the precise determination of two-body interactions. This is summarized in Fig.~\ref{fig:twobody}. Since these results correspond to the physical point, they can be compared to dispersive analyses for $\pi^+\pi^+$ and $\pi^+ K^+$~\cite{Garcia-Martin:2011iqs,Pelaez:2020gnd}. We generally find good agreement with these analyses and, notably, achieve even lower statistical uncertainties in some cases. Additionally, we obtain some constraints on higher partial waves, although only in the $K^+K^+$ system does the $d$-wave deviate from zero by more than one sigma.

\begin{figure}[th!]
     \centering
     \subfloat[\label{fig:K0}]{%
     \includegraphics[width=0.49\textwidth]{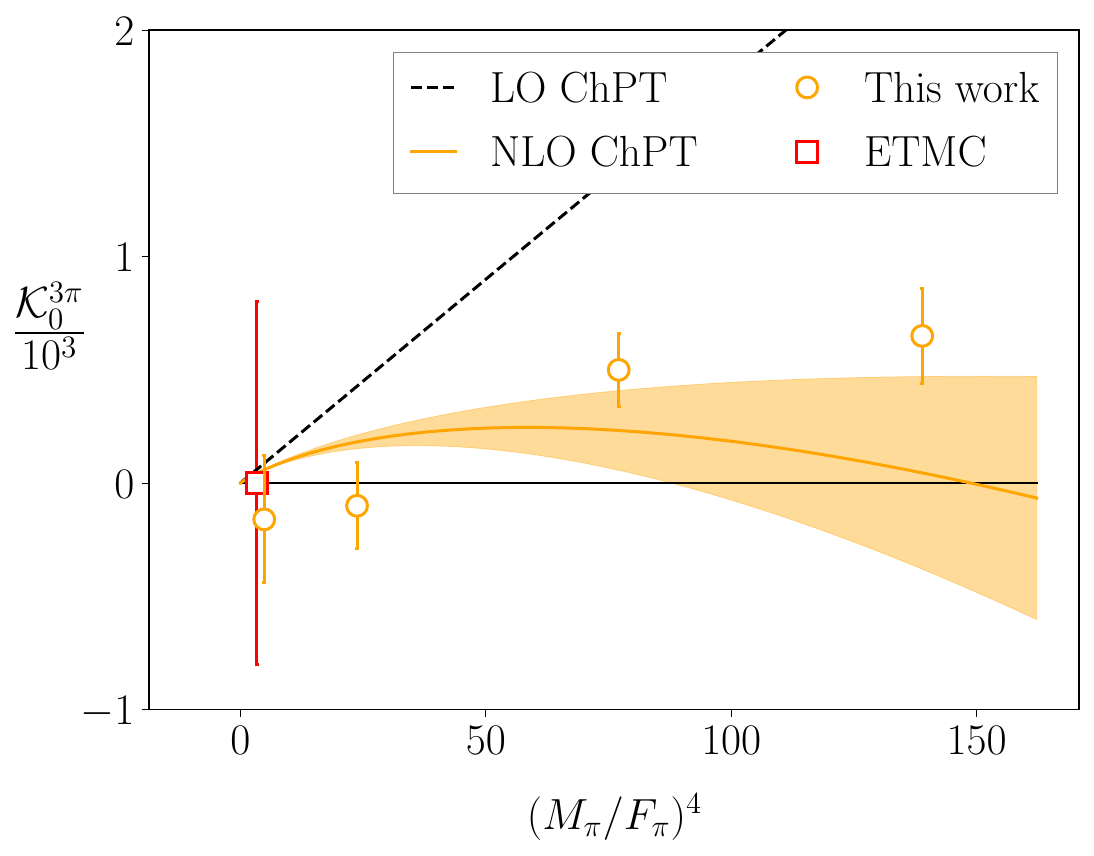}
    }
    \hfill
    \subfloat[\label{fig:K1}]{%
     \includegraphics[width=0.49\textwidth]{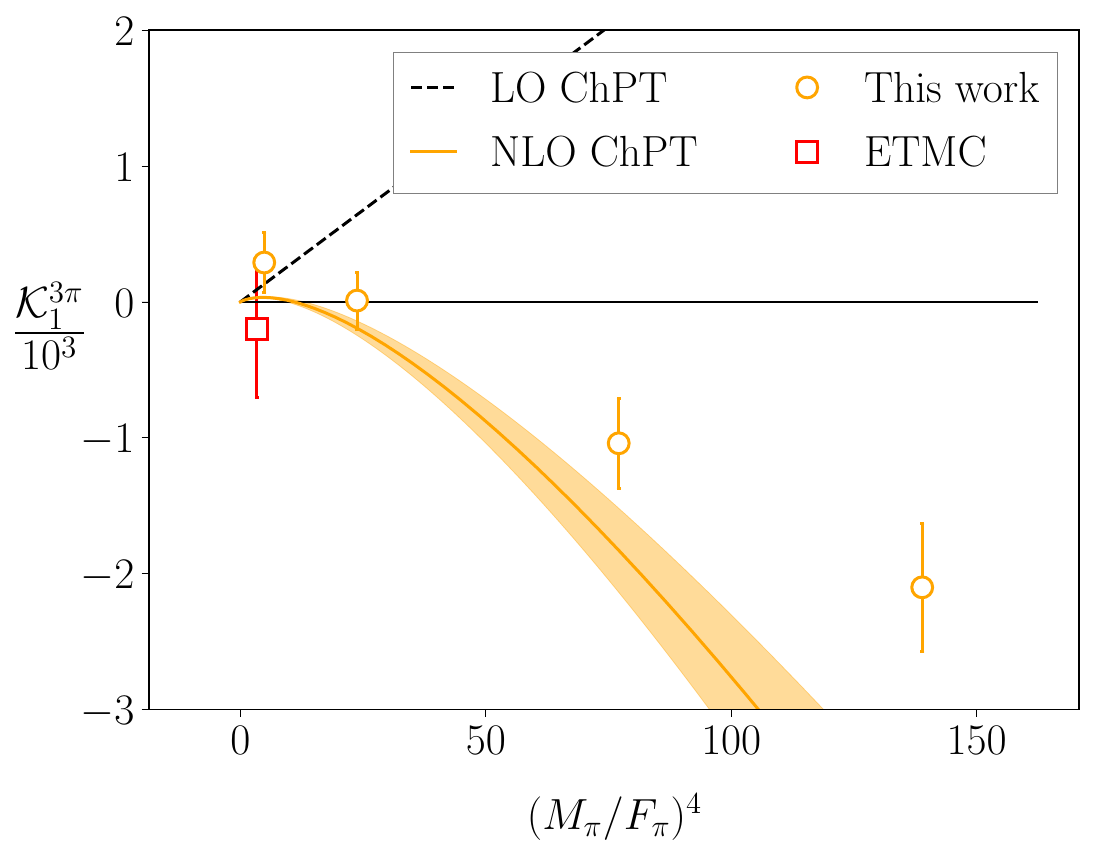}
    }
    \caption{Results for the lowest two coefficients, $\mathcal{K}_0$ (left) and $\mathcal{K}_1$ (right), in the threshold expansion of the $3\pi^+$ $\mathcal{K}_{\rm df,3}$. Orange circles correspond to our latest results~\cite{Dawid:2025doq,Dawid:2025zxc,Blanton:2021llb}, while red squares represent results from Ref.~\cite{Fischer:2020jzp}. Leading-order (LO) and next-to-leading-order (NLO) ChPT predictions are also shown~\cite{Baeza-Ballesteros:2023ljl}. The error bands reflect the uncertainties in the low-energy constants (LECs), as detailed in Ref.~\cite{Baeza-Ballesteros:2023ljl}.
 }
    \label{fig:K01}
\end{figure}

From the fit results, we can examine the values of the K-matrix in the $3\pi^+$ system, for which Chiral Perturbation Theory (ChPT) predictions are available at leading and next-to-leading orders (LO and NLO)~\cite{Bijnens:2021hpq,Bijnens:2022zsq,Baeza-Ballesteros:2023ljl,Baeza-Ballesteros:2024mii}. This comparison is meaningful despite the scheme dependence of the K matrix, as the ChPT calculation can be performed within the same scheme. Figure~\ref{fig:K01} presents the results for the four ensembles used in this study, including a physical-point result from the ETMC collaboration~\cite{Fischer:2020jzp} and the ChPT predictions. For the NLO ChPT curve, phenomenologically determined low-energy constants (LECs) are used as inputs, see Ref.~\cite{Baeza-Ballesteros:2023ljl}. As shown in the figure, qualitative agreement is found between the lattice QCD results and NLO ChPT. A striking feature, however, is the significant shift from LO to NLO. Whether this poor convergence persists at higher orders remains an open question, requiring further higher-order calculations.

\begin{figure}[th!]
     \centering
    \includegraphics[width=0.51\textwidth]{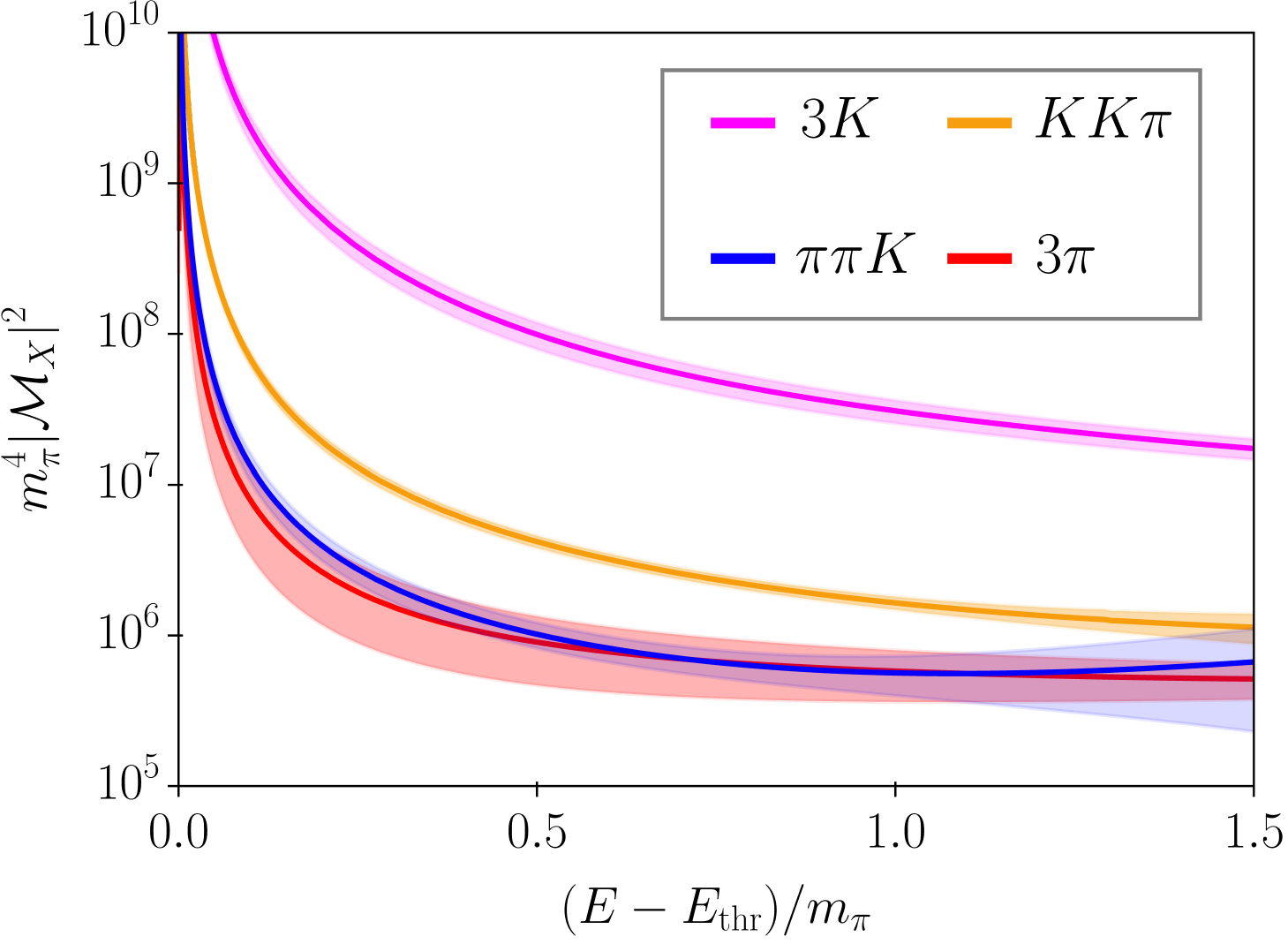}
    \caption{ $J^P=0^-$ three-hadron scattering amplitude as function of the energy difference to threshold for systems of pions and kaons at maximal isospin. The incoming and outgoing particles lay on an equilateral kinematic configuration, so that fixing the energy completely defines all kinematic variables. 
    } \label{fig:M3}
\end{figure}

After computing the two- and three-meson K matrices from the lattice QCD spectrum, the corresponding scattering amplitudes are obtained by solving integral equations projected onto definite angular momentum and parity, \(J^P\). However, visualizing these results is challenging, as three-meson scattering amplitudes depend on eight kinematic variables after accounting for Poincaré invariance. To display the results in a two-dimensional plot, we choose kinematic configurations of incoming and outgoing momenta. One such choice is the equilateral configuration, where the momenta of the incoming and outgoing particles form equilateral triangles. Then, the scattering amplitude can be plotted as a function of energy. Figure~\ref{fig:M3} presents results for systems of three mesons composed of either $\pi^+$ or $K^+$. Two key features are visible. First, all amplitudes diverge at threshold, due to one-particle exchange processes. Second, the $3K^+$ amplitude is significantly larger than the others, consistent with the naive chiral expectation that kaon interactions are stronger due to the heavier strange-quark mass.

\begin{figure}[h!]
     \centering
    \includegraphics[width=0.59\textwidth]{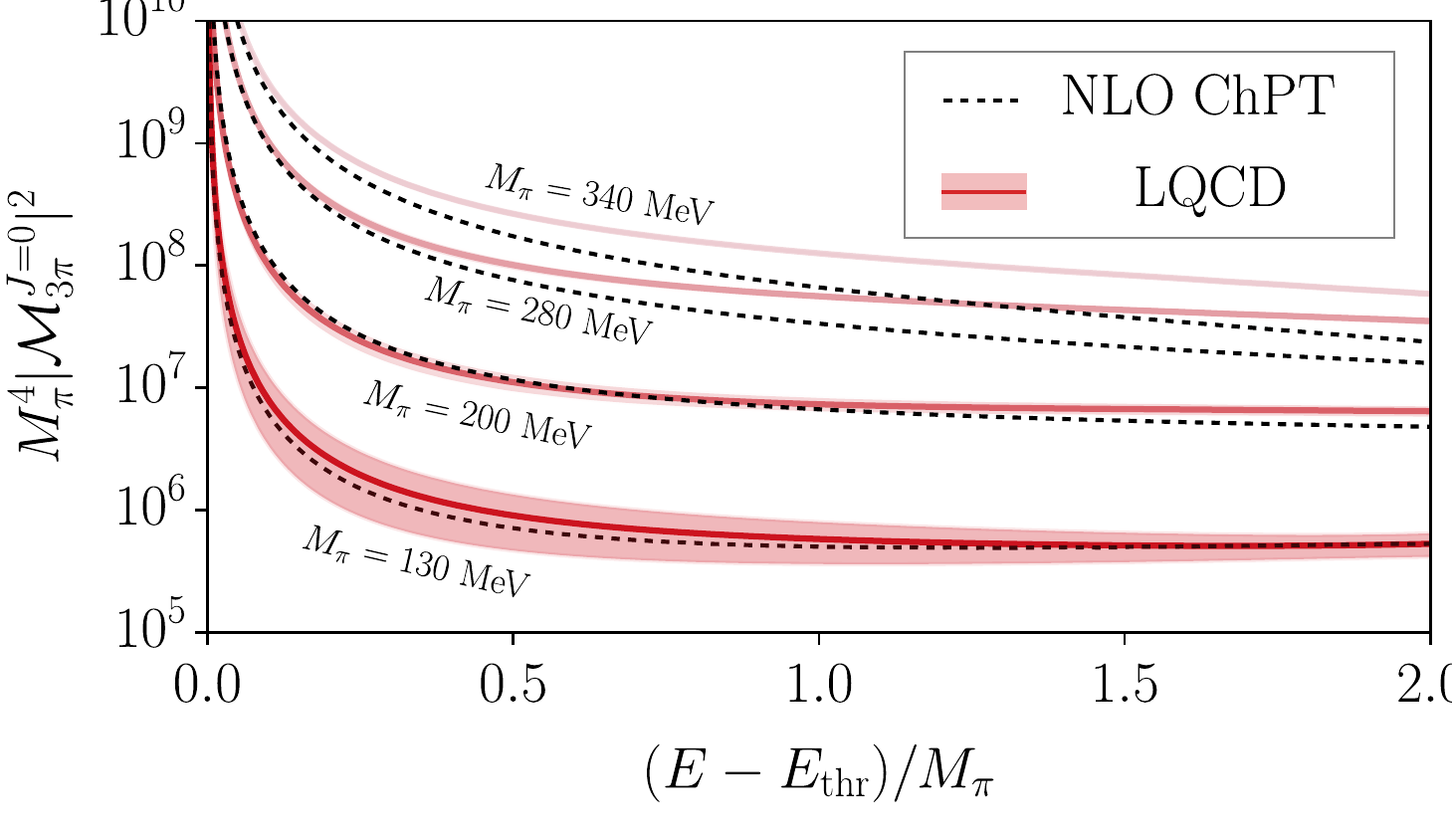}
    \caption{ Three-pion scattering amplitude at maximal isospin for four different ensembles with quark masses ranging from approximately physical to around 340 MeV. The ChPT predictions~\cite{Bijnens:2021hpq,Baeza-Ballesteros:2023ljl} are also shown.
    } \label{fig:chiralM3}
\end{figure}

To conclude the discussion of these systems, we investigate the chiral dependence of the $3\pi^+$ amplitude. Figure~\ref{fig:chiralM3} presents the results for the 
$J^P=0^-$ amplitude on the equilateral kinematic configuration and on all
four ensembles. As can be seen, pion interactions become stronger with increasing pion mass. The full amplitude can also be compared to NLO ChPT predictions, showing good agreement at the physical point. However, the agreement deteriorates at heavier pion masses and higher energies, which is consistent with the power counting of the chiral expansion.

\clearpage
\section{Towards a three-body description of the doubly-charmed tetraquark}

Motivated by the discovery of the doubly-charmed tetraquark, $T_{\rm cc}$, by LHCb via its three-hadron decay mode $T_{\rm cc} \to DD\pi$~\cite{LHCb:2021vvq,LHCb:2021auc}, there has been a growing effort to predict its properties using lattice QCD~\cite{Padmanath:2022cvl,Lyu:2023xro,Chen:2022vpo,Ortiz-Pacheco:2023ble,Collins:2024sfi,Whyte:2024ihh,Stump:2024lqx,Vujmilovic:2024snz,Meng:2024kkp} and phenomenological analyses~\cite{Albaladejo:2021vln,Du:2021zzh,Achasov:2022onn,Wang:2023iaz,Zhang:2024dth,Du:2025vkm,Abolnikov:2024key}. Since its decay involves three hadrons, a rigorous theoretical treatment requires a three-body formalism at physical quark masses.

\begin{figure}[th!]
     \centering
     \subfloat[\label{fig:tc}]{%
     \includegraphics[width=0.4\textwidth]{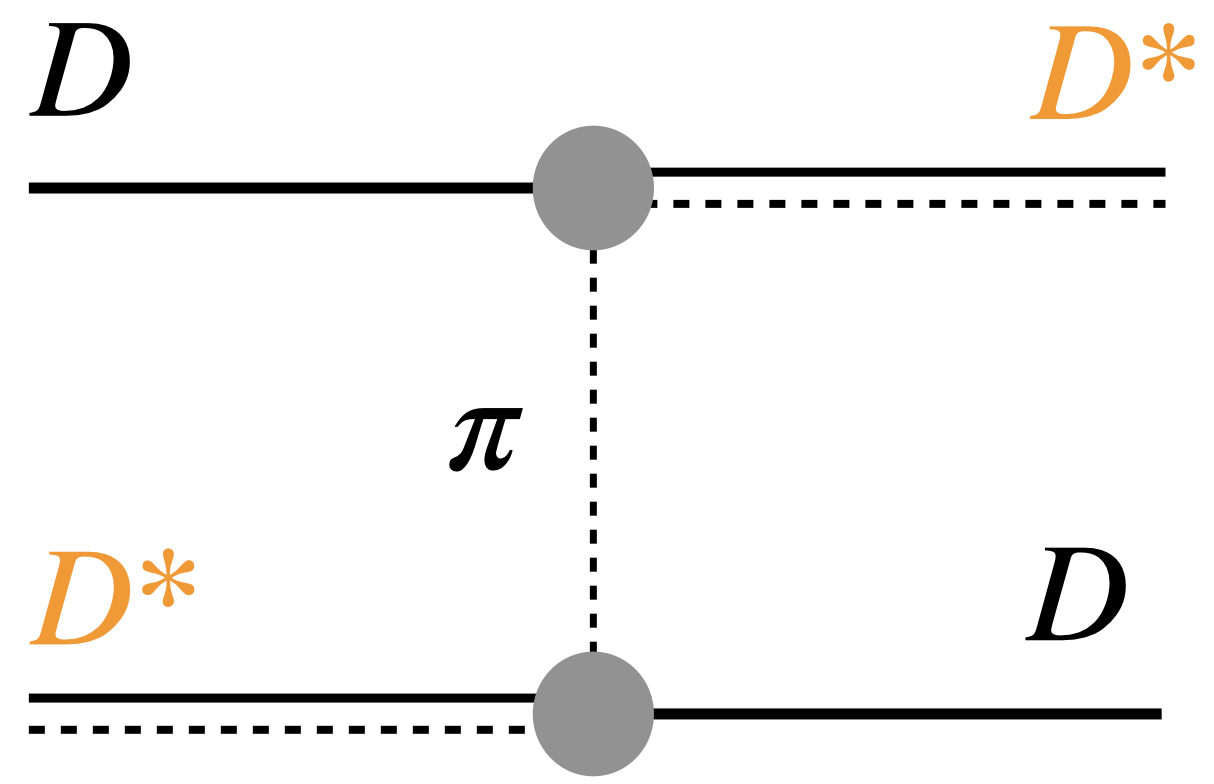}
    }
    \hfill
    \subfloat[\label{fig:gd}]{%
     \includegraphics[width=0.49\textwidth]{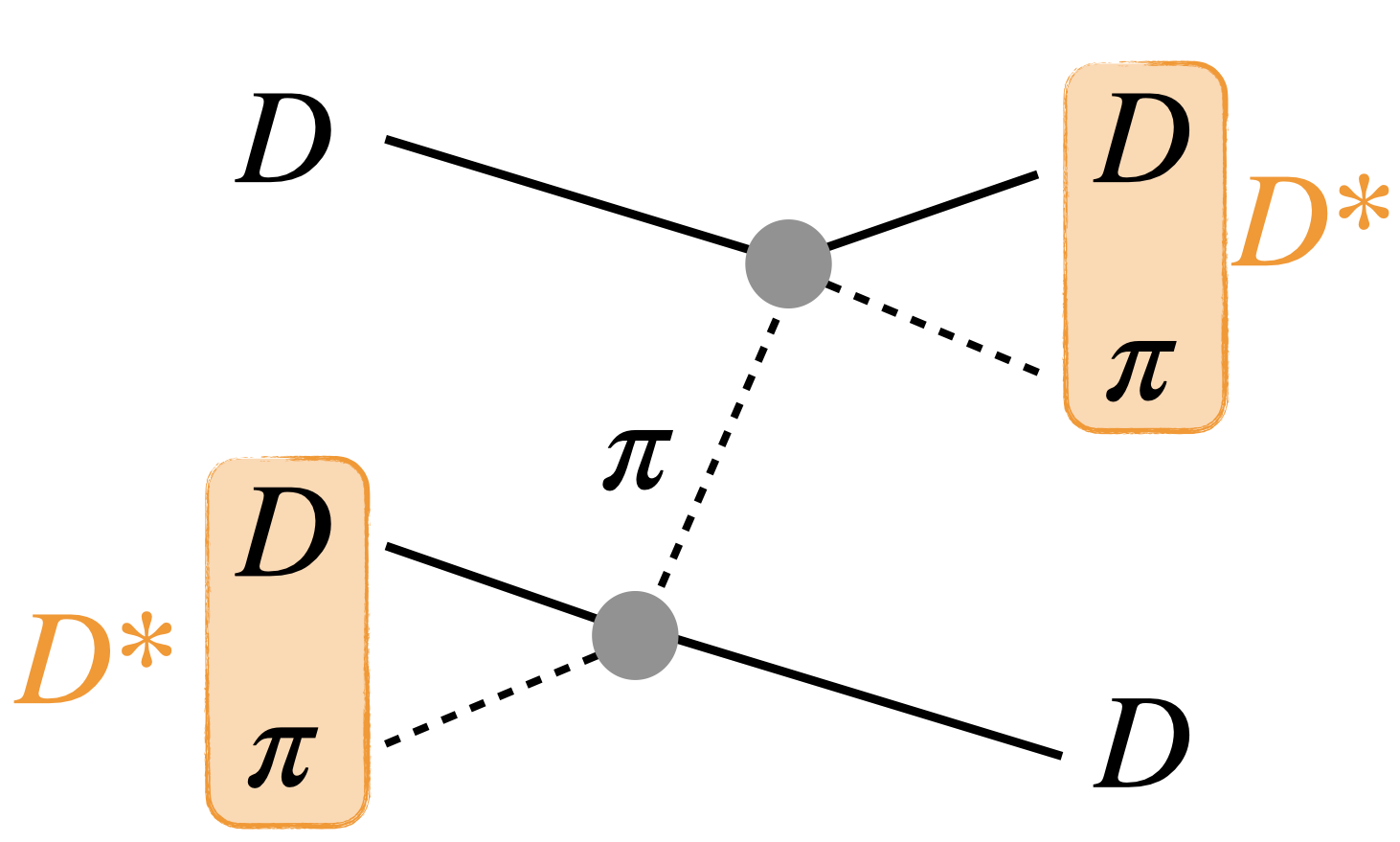}
    }
    \caption{ Left: one-pion exchange diagram in $DD^*$ scattering responsible for the left-hand cut. Right: equivalent diagram in the context of $DD\pi$ scattering, where the $D^*$ meson is viewed as pole in the $D\pi$ amplitude.  }
    \label{fig:lhc}
\end{figure}

In many lattice QCD studies, where the light quark mass is heavier than physical, the $D^*$ meson becomes stable. In such cases, the extraction of $T_{\rm cc}$ properties can be formulated in terms of two-body $DD^*$ scattering. However, the $DD^*$ amplitude features a left-hand cut, which complicates the determination of the pole position and cannot be neglected---see Fig.~\ref{fig:tc}. To address this issue, several approaches have been proposed, including explicitly incorporating pion exchange in $DD^*$ scattering~\cite{Meng:2023bmz,Du:2023hlu,Bubna:2024izx,Raposo:2023oru,Raposo:2025dkb,Du:2021zzh} (see also Ref.~\cite{HansenCD}).

Our proposal, however, is to treat the system as a genuine three-meson system when the $D^*$ meson is stable, using the formalism derived in Ref.~\cite{Hansen:2024ffk}. In this approach, inspired by Refs.~\cite{Jackura:2020bsk,Dawid:2023jrj}, we incorporate the $D^*$ meson as a bound-state pole in the $I=1/2$ $D\pi$ scattering amplitude. This allows the finite-volume energies of the $DD^*$ system to be computed using the three-body quantization condition. This approach is motivated by the explicit inclusion of the diagram in Fig.~\ref{fig:gd} in the three-body formalism. An important advantage of this method over two-body treatments is that it enables a smooth connection between the regimes where the $D^*$ is stable and unstable, both in finite and infinite volume. However, it is also qualitatively more involved, as a full analysis requires input for both the $DD$ and $D\pi$ scattering amplitudes.

In a recent study~\cite{Dawid:2024dgy}, we applied the three-body formalism to analyze the finite-volume energies from Ref.~\cite{Padmanath:2022cvl}. Our analysis requires as input the $s$-wave $I=1$ $DD$ scattering amplitude, the $I=1/2$ $D\pi$ scattering amplitude in both $s$- and $p$-waves, and a parametrization of the three-particle K matrix. For the three-body K matrix, we adopt a parametrization based on the first term in the threshold expansion that couples to $J^P=1^+$:
\begin{equation}
    \mathcal{K}_{\rm df,3} = \mathcal{K}_E (p_\pi - k_\pi)^2,
\end{equation}
where $\mathcal{K}_E$ is a constant (i.e., an adjustable parameter), and $p_\pi$ and $k_\pi$ denote the momenta of the incoming/outgoing pions in $DD\pi$ scattering. However, since Ref.~\cite{Padmanath:2022cvl} does not provide $DD$ or $D\pi$ energy levels, we use input inspired by other lattice QCD results~\cite{Yan:2024yuq} and effective field theories~\cite{Belyaev:1994zk,BaBar:2013zgp} for the two-body amplitudes. In practice, we neglect $DD$ scattering (as it is weakly interacting), and employ parametrizations for the $D\pi$ amplitude that incorporate the $D^*$ meson as a stable particle in the $p$-wave, and the $D_0^*$ resonance in the $s$-wave.

\begin{figure}[th!]
     \centering
    \includegraphics[width=0.7\textwidth]{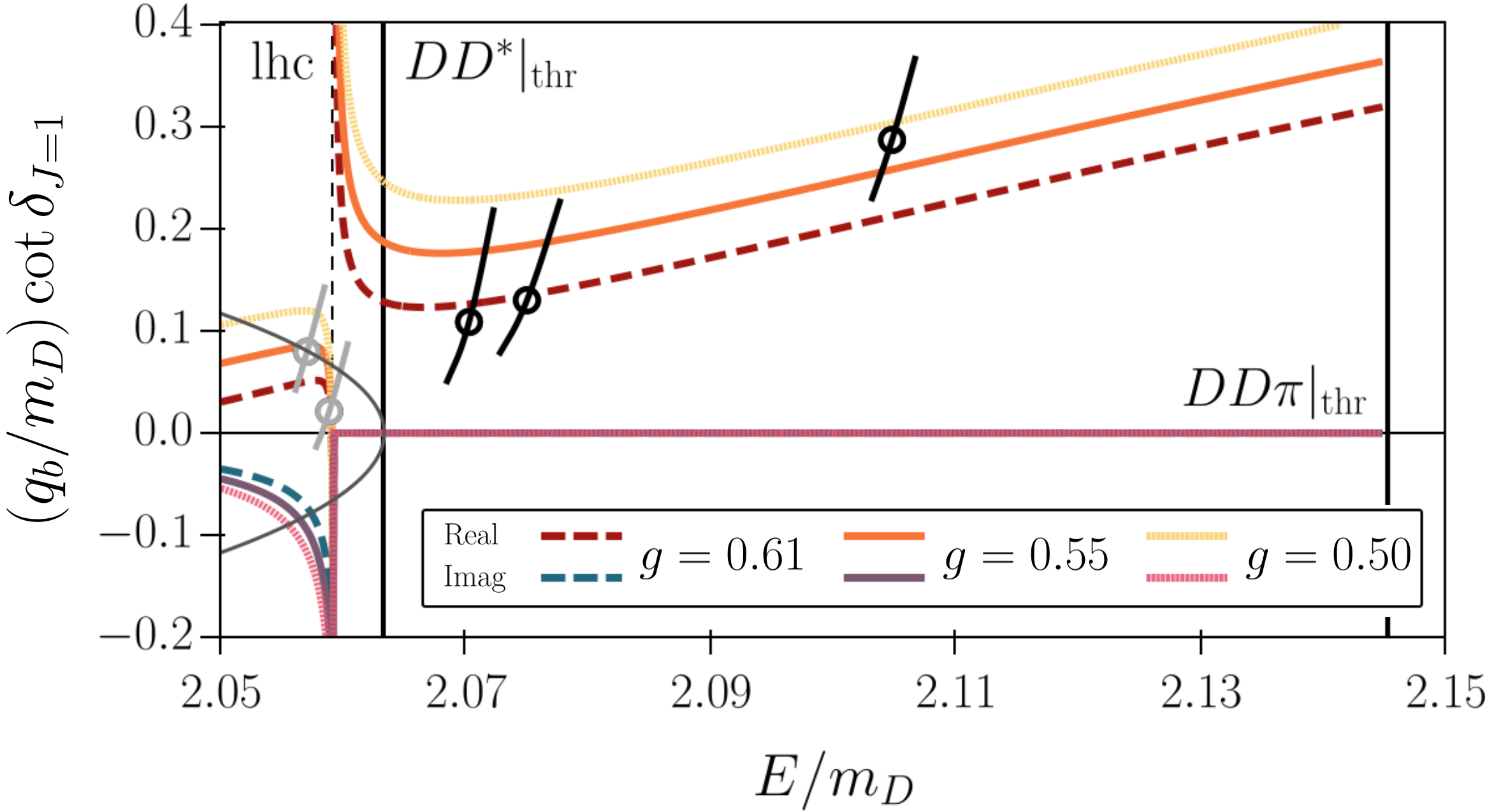}
    \caption{ $DD^*$ $s$-wave phase shift as a function of energy. Points with error bars come from the energies in Ref.~\cite{Padmanath:2022cvl} using the two-body formalism. The ones in grey indicate that the left-hand cut invalidates the applicability of the original two-body approach. Three curves are shown for different choices of the $DD^*\pi$ coupling in the $D\pi$ amplitude, $g$. In all cases, $m^2_D\mathcal K_E = 1.9\cdot10^5$. As can be seen, the phase shift becomes complex on the left-hand cut.
    } \label{fig:phaseDDs}
\end{figure}

The results of this analysis are shown in Fig.~\ref{fig:phaseDDs}, where we plot the phase shift in the "\( q\cot\delta \)" form as a function of energy. The values of $\mathcal{K}_E$ have been tuned such that the resulting curves reproduce the phase shifts extracted from the above-threshold energies of Ref.~\cite{Padmanath:2022cvl} using the two-body formalism. As evident from the plot, the $q\cot\delta$ function develops a non-zero imaginary part on the left-hand cut---a sign of the non-analyticity at the branch point. The existence of this imaginary component at real energies rules out the presence of bound or virtual state poles for real energies on the left-hand cut for this choice of parameters.

\begin{figure}[th!]
     \centering
    \includegraphics[width=0.7\textwidth]{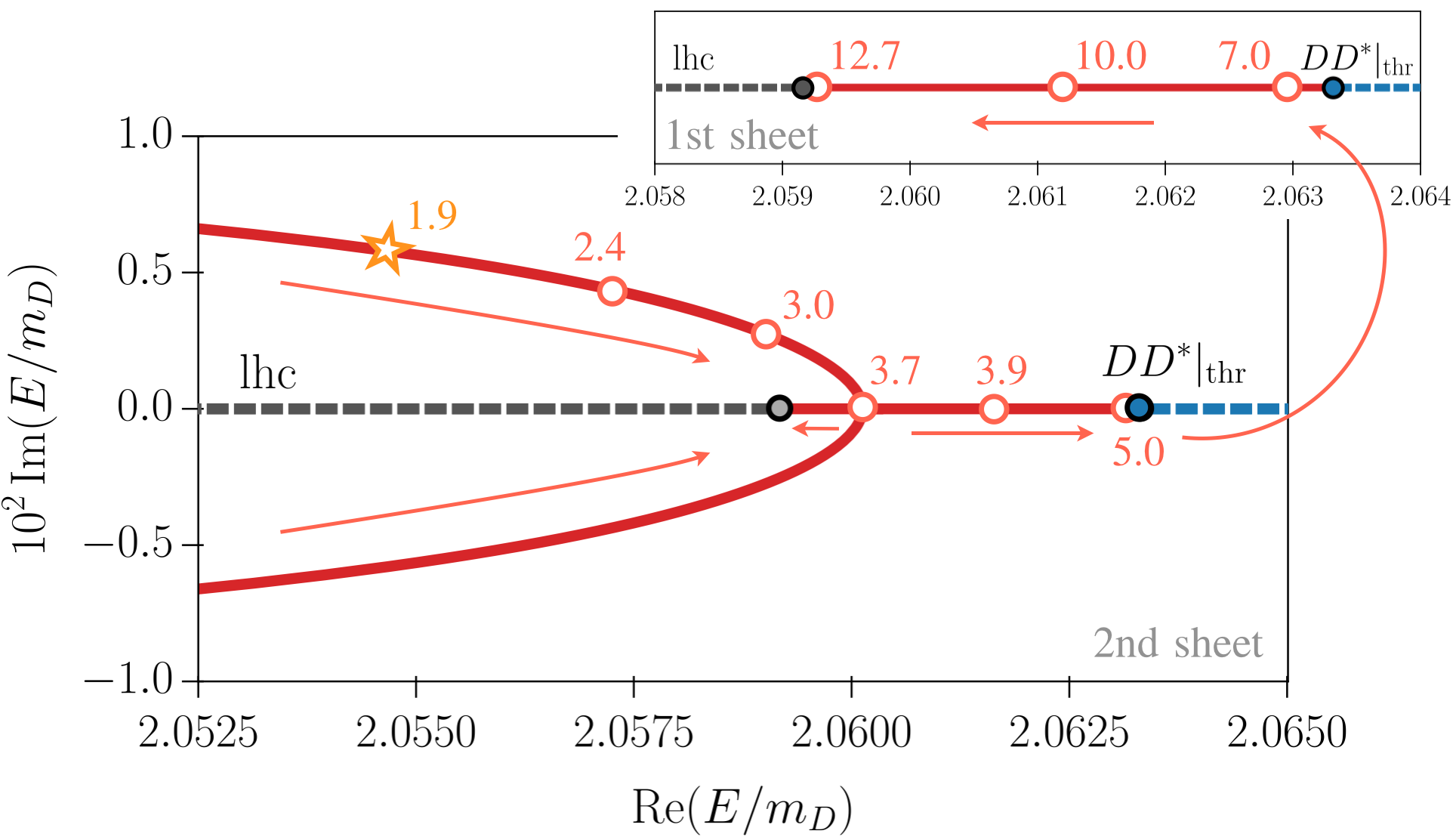}
    \caption{ $T_{\rm cc}$ pole position in the complex plane as a function of the parameter in the three -body K matrix. The numbers of the plot indicate the value of  $\mathcal K_E$ in units  $m_D^2 \mathcal K_E / 10^5$. The one matching the setup in Fig.~\ref{fig:phaseDDs} is marked with a star. The positions of the left-hand cut (lhc) and of the threshold (thr) are also displayed.
    } \label{fig:polepos}
\end{figure}

The integral equations also allow for an analytic continuation of the $DD^*$ scattering amplitude into the complex energy plane, enabling the search for poles. In Fig.~\ref{fig:polepos}, we display the location of the $T_{\rm cc}$ pole in the complex plane as a function of $\mathcal{K}_E$, keeping the two-body parameters fixed. At the value of $\mathcal{K}_E$ that best describes the finite-volume data (corresponding to Fig.~\ref{fig:phaseDDs}), we find two poles with non-zero imaginary parts. These poles, which have been observed in other analyses~\cite{Abolnikov:2024key}, correspond to a so-called "subthreshold resonance." As $\mathcal{K}_E$ is increased, the poles move to higher energies and approach the real axis. Eventually, they transition into a pair of virtual state poles, and with further increase of $\mathcal{K}_E$, one of them becomes a bound-state pole.

\begin{figure}[th!]
     \centering
    \includegraphics[width=\textwidth]{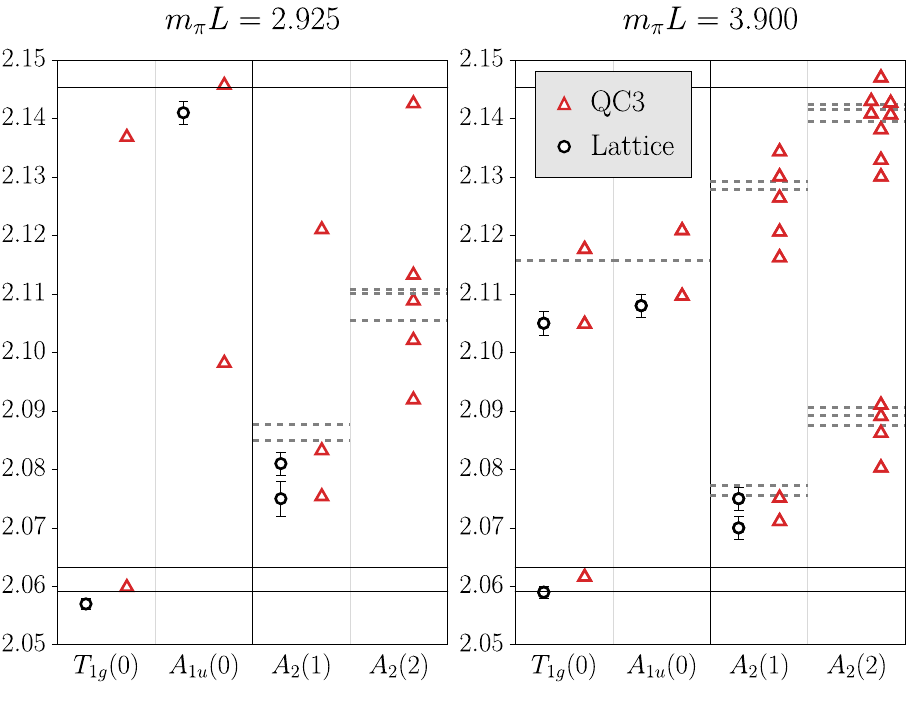}
    \caption{ Finite-volume spectrum of the $DD^*$ system in several irreps and frames. The results from Ref.~\cite{Padmanath:2022cvl} are shown as black circles. The energies predicted by the QC3, using the same setup as Fig.~\ref{fig:phaseDDs}, are shown as red triangles.  Non-interacting levels are marked as horizontal grey dashed lines.  } \label{fig:DDsspectrum}
\end{figure}

To conclude this topic, we perform a finite-volume analysis of the $DD^*$ spectrum using the three-body quantization condition. As input, we use the same parametrizations of the K matrices employed for the central curve in Fig.~\ref{fig:phaseDDs}. The resulting  QC3 spectra for several irreps and moving frames are shown in Fig.~\ref{fig:DDsspectrum}, where they are compared to the finite-volume energies from Ref.~\cite{Padmanath:2022cvl}. Overall, we find good agreement. However, a notable discrepancy is present in the $A_{1u}(0)$ irrep, where the QC3 predicts an energy level in the smaller volume that was not observed in the lattice QCD calculation. We interpret this level qualitatively as a ``$DD\pi$''-like state, shifted downward by the attractive threshold interactions in the $s$-wave $D\pi$ channel due to the proximity of the $D_0^*$ resonance. We suspect that this level was missed in the lattice calculation due to the absence of $DD\pi$-type interpolating operators. This further highlights the importance of a genuine three-body analysis even in cases where the $D^*$ meson is stable.

\section{Conclusion and Outlook}

In this contribution, I have highlighted some recent advances in three-hadron spectroscopy. Specifically, I discussed recent calculations of the three-meson scattering amplitude at physical quark masses and our three-body study of the doubly-charmed tetraquark. 

The first topic marks a significant milestone in three-body studies, demonstrating that lattice QCD techniques are now capable of physical-point calculations and that subleading three-body effects can be resolved in certain systems~\cite{Dawid:2025doq,Dawid:2025zxc}.  

For the second topic, we have fully developed a strategy to perform lattice QCD calculations of the \(T_{\rm cc}\), incorporating both three-body effects and left-hand cuts. Our results successfully reproduce the existing lattice QCD spectrum from Ref.~\cite{Padmanath:2022cvl}, and show qualitative agreement with other approaches to \(DD^*\) scattering. However, we have argued that it is crucial to perform a three-body analysis even if the $D^*$ meson is bound, as the proximity of the $DD\pi$ threshold can have effects in energy determinations, as well as finite-volume effects.

Progress in this subfield has been remarkable over the past decade. Formalism exists to deal with complicated three-hadron systems, such as coupled-channel three-meson systems~\cite{Draper:2024qeh}. However, several open challenges remain in three-particle spectroscopy. First, many existing theoretical tools still need to be applied to lattice QCD systems, specifically considering three-hadron resonances~\cite{Hansen:2020zhy}, and even three baryons~\cite{Draper:2023xvu} or electroweak decays~\cite{Muller:2020wjo,Hansen:2021ofl,Pang:2023jri,Muller:2022oyw}.  Beyond numerical applications, further advancements in formalism are essential, particularly in addressing coupled-channel systems that include baryons. On a more theoretical level, the analytic continuation of three-hadron scattering amplitudes remains a challenging problem, especially due to the smooth coupling inherent in the RFT formalism~\cite{Dawid:2023jrj}.

The study of hadron properties and interactions remains a significant challenge within the Standard Model, and advances in three-hadron spectroscopy are providing new theoretical perspectives and valuable phenomenological insights.

\section*{Acknowledgements}

I would like to thank the organizers of CD24 for giving the opportunity to give this plenary talk. Moreover, I would like to thank Vadim Baru, Sara Collins, Jeremy Green, Meng Lin Du, and Saša Prelovšek for discussions directly related to my presentation at CD24.

Special thanks to my collaborators Sebastian Dawid, Zack Draper, Colin Morningstar, Andrew Hanlon, Ben H\"orz, Steve Sharpe and Sarah Skinner for an enjoyable collaboration during the research presented here. I am also grateful to Zack Draper and Steve Sharpe for their comments on this manuscript.

Data from Ref.~\cite{Padmanath:2022cvl} were kindly provided by Saša Prelovšek and Madanagopalan Padmanath.

In Refs.~\cite{Dawid:2025doq,Dawid:2025zxc}, computations were performed in part in Frontera at the Texas Advanced Computing Center (U.S. NSF, award OAC-1818253), in Hawk at the High Performance Computing Center in Stuttgart, and in “Mogon II" at JGU Mainz.

\bibliographystyle{JHEP}
\bibliography{cited_refs}   

\providecommand{\href}[2]{#2}\begingroup\raggedright\begin{thebibliography}{100}

\bibitem{LHCb-FIGURE-2021-001}
{\scshape LHCb} collaboration, \emph{{List of hadrons observed at the LHC}}, .

\bibitem{Liu:2023hhl}
Z.~Liu and R.E.~Mitchell, \emph{{New hadrons discovered at BESIII}},
  \href{https://doi.org/10.1016/j.scib.2023.08.025}{\emph{Sci. Bull.}
  {\bfseries 68} (2023) 2148}
  [\href{https://arxiv.org/abs/2310.09465}{{\ttfamily 2310.09465}}].

\bibitem{Chen:2022asf}
H.-X.~Chen, W.~Chen, X.~Liu, Y.-R.~Liu and S.-L.~Zhu, \emph{{An updated review
  of the new hadron states}},
  \href{https://doi.org/10.1088/1361-6633/aca3b6}{\emph{Rept. Prog. Phys.}
  {\bfseries 86} (2023) 026201}
  [\href{https://arxiv.org/abs/2204.02649}{{\ttfamily 2204.02649}}].

\bibitem{LHCb:2021vvq}
{\scshape LHCb} collaboration, \emph{{Observation of an exotic narrow doubly
  charmed tetraquark}},
  \href{https://doi.org/10.1038/s41567-022-01614-y}{\emph{Nature Phys.}
  {\bfseries 18} (2022) 751}
  [\href{https://arxiv.org/abs/2109.01038}{{\ttfamily 2109.01038}}].

\bibitem{LHCb:2021auc}
{\scshape LHCb} collaboration, \emph{{Study of the doubly charmed tetraquark
  $T_{cc}^{+}$}},
  \href{https://doi.org/10.1038/s41467-022-30206-w}{\emph{Nature Commun.}
  {\bfseries 13} (2022) 3351}
  [\href{https://arxiv.org/abs/2109.01056}{{\ttfamily 2109.01056}}].

\bibitem{Burkert:2017djo}
V.D.~Burkert and C.D.~Roberts, \emph{{Colloquium : Roper resonance: Toward a
  solution to the fifty year puzzle}},
  \href{https://doi.org/10.1103/RevModPhys.91.011003}{\emph{Rev. Mod. Phys.}
  {\bfseries 91} (2019) 011003}
  [\href{https://arxiv.org/abs/1710.02549}{{\ttfamily 1710.02549}}].

\bibitem{Mai:2022eur}
M.~Mai, U.-G.~Mei\ss{}ner and C.~Urbach, \emph{{Towards a theory of hadron
  resonances}},
  \href{https://doi.org/10.1016/j.physrep.2022.11.005}{\emph{Phys. Rept.}
  {\bfseries 1001} (2023) 1}
  [\href{https://arxiv.org/abs/2206.01477}{{\ttfamily 2206.01477}}].

\bibitem{Hammer:2012id}
H.-W.~Hammer, A.~Nogga and A.~Schwenk, \emph{{Three-body forces: From cold
  atoms to nuclei}},
  \href{https://doi.org/10.1103/RevModPhys.85.197}{\emph{Rev. Mod. Phys.}
  {\bfseries 85} (2013) 197} [\href{https://arxiv.org/abs/1210.4273}{{\ttfamily
  1210.4273}}].

\bibitem{Bombaci:2016xzl}
I.~Bombaci, \emph{{The Hyperon Puzzle in Neutron Stars}},
  \href{https://doi.org/10.7566/JPSCP.17.101002}{\emph{JPS Conf. Proc.}
  {\bfseries 17} (2017) 101002}
  [\href{https://arxiv.org/abs/1601.05339}{{\ttfamily 1601.05339}}].

\bibitem{Gerstung:2020ktv}
D.~Gerstung, N.~Kaiser and W.~Weise, \emph{{Hyperon\textendash{}nucleon
  three-body forces and strangeness in neutron stars}},
  \href{https://doi.org/10.1140/epja/s10050-020-00180-2}{\emph{Eur. Phys. J. A}
  {\bfseries 56} (2020) 175}
  [\href{https://arxiv.org/abs/2001.10563}{{\ttfamily 2001.10563}}].

\bibitem{NA482:2007ucr}
{\scshape NA48/2} collaboration, \emph{{Search for direct CP violating charge
  asymmetries in K+- ---\ensuremath{>} pi+- pi+ pi- and K+- ---\ensuremath{>}
  pi+- pi0 pi0 decays}},
  \href{https://doi.org/10.1140/epjc/s10052-007-0456-7}{\emph{Eur. Phys. J. C}
  {\bfseries 52} (2007) 875} [\href{https://arxiv.org/abs/0707.0697}{{\ttfamily
  0707.0697}}].

\bibitem{NA482:2010gwp}
{\scshape NA48/2} collaboration, \emph{{Empirical parameterization of the K+-
  -\ensuremath{>} pi+- pi0 pi0 decay Dalitz plot}},
  \href{https://doi.org/10.1016/j.physletb.2010.02.036}{\emph{Phys. Lett. B}
  {\bfseries 686} (2010) 101}
  [\href{https://arxiv.org/abs/1004.1005}{{\ttfamily 1004.1005}}].

\bibitem{Detmold:2008gh}
W.~Detmold and M.J.~Savage, \emph{{The Energy of n Identical Bosons in a Finite
  Volume at O($L^{-7})$}},
  \href{https://doi.org/10.1103/PhysRevD.77.057502}{\emph{Phys. Rev. D}
  {\bfseries 77} (2008) 057502}
  [\href{https://arxiv.org/abs/0801.0763}{{\ttfamily 0801.0763}}].

\bibitem{Beane:2007qr}
S.R.~Beane, W.~Detmold and M.J.~Savage, \emph{{n-Boson Energies at Finite
  Volume and Three-Boson Interactions}},
  \href{https://doi.org/10.1103/PhysRevD.76.074507}{\emph{Phys. Rev. D}
  {\bfseries 76} (2007) 074507}
  [\href{https://arxiv.org/abs/0707.1670}{{\ttfamily 0707.1670}}].

\bibitem{Briceno:2012rv}
R.A.~Briceno and Z.~Davoudi, \emph{{Three-particle scattering amplitudes from a
  finite volume formalism}},
  \href{https://doi.org/10.1103/PhysRevD.87.094507}{\emph{Phys. Rev.}
  {\bfseries D87} (2013) 094507}
  [\href{https://arxiv.org/abs/1212.3398}{{\ttfamily 1212.3398}}].

\bibitem{Polejaeva:2012ut}
K.~Polejaeva and A.~Rusetsky, \emph{{Three particles in a finite volume}},
  \href{https://doi.org/10.1140/epja/i2012-12067-8}{\emph{Eur. Phys. J.}
  {\bfseries A48} (2012) 67} [\href{https://arxiv.org/abs/1203.1241}{{\ttfamily
  1203.1241}}].

\bibitem{Hansen:2014eka}
M.T.~Hansen and S.R.~Sharpe, \emph{{Relativistic, model-independent,
  three-particle quantization condition}},
  \href{https://doi.org/10.1103/PhysRevD.90.116003}{\emph{Phys. Rev. D}
  {\bfseries 90} (2014) 116003}
  [\href{https://arxiv.org/abs/1408.5933}{{\ttfamily 1408.5933}}].

\bibitem{Hansen:2015zga}
M.T.~Hansen and S.R.~Sharpe, \emph{{Expressing the three-particle finite-volume
  spectrum in terms of the three-to-three scattering amplitude}},
  \href{https://doi.org/10.1103/PhysRevD.92.114509}{\emph{Phys. Rev.}
  {\bfseries D92} (2015) 114509}
  [\href{https://arxiv.org/abs/1504.04248}{{\ttfamily 1504.04248}}].

\bibitem{Briceno:2017tce}
R.A.~{Brice{\~n}o}, M.T.~Hansen and S.R.~Sharpe, \emph{{Relating the
  finite-volume spectrum and the two-and-three-particle $S$ matrix for
  relativistic systems of identical scalar particles}},
  \href{https://doi.org/10.1103/PhysRevD.95.074510}{\emph{Phys. Rev.}
  {\bfseries D95} (2017) 074510}
  [\href{https://arxiv.org/abs/1701.07465}{{\ttfamily 1701.07465}}].

\bibitem{Hammer:2017uqm}
H.-W.~Hammer, J.-Y.~Pang and A.~Rusetsky, \emph{{Three-particle quantization
  condition in a finite volume: 1. The role of the three-particle force}},
  \href{https://doi.org/10.1007/JHEP09(2017)109}{\emph{JHEP} {\bfseries 09}
  (2017) 109} [\href{https://arxiv.org/abs/1706.07700}{{\ttfamily
  1706.07700}}].

\bibitem{Hammer:2017kms}
H.W.~Hammer, J.Y.~Pang and A.~Rusetsky, \emph{{Three particle quantization
  condition in a finite volume: 2. general formalism and the analysis of
  data}}, \href{https://doi.org/10.1007/JHEP10(2017)115}{\emph{JHEP} {\bfseries
  10} (2017) 115} [\href{https://arxiv.org/abs/1707.02176}{{\ttfamily
  1707.02176}}].

\bibitem{Mai:2017bge}
M.~Mai and M.~D\"oring, \emph{{Three-body Unitarity in the Finite Volume}},
  \href{https://doi.org/10.1140/epja/i2017-12440-1}{\emph{Eur. Phys. J. A}
  {\bfseries 53} (2017) 240}
  [\href{https://arxiv.org/abs/1709.08222}{{\ttfamily 1709.08222}}].

\bibitem{Briceno:2018aml}
R.A.~{Brice\~{n}o}, M.T.~Hansen and S.R.~Sharpe, \emph{{Three-particle systems
  with resonant subprocesses in a finite volume}},
  \href{https://arxiv.org/abs/1810.01429}{{\ttfamily 1810.01429}}.

\bibitem{Briceno:2018mlh}
R.A.~Brice\~no, M.T.~Hansen and S.R.~Sharpe, \emph{{Numerical study of the
  relativistic three-body quantization condition in the isotropic
  approximation}},
  \href{https://doi.org/10.1103/PhysRevD.98.014506}{\emph{Phys. Rev. D}
  {\bfseries 98} (2018) 014506}
  [\href{https://arxiv.org/abs/1803.04169}{{\ttfamily 1803.04169}}].

\bibitem{Pang:2019dfe}
J.-Y.~Pang, J.-J.~Wu, H.W.~Hammer, U.-G.~Mei\ss{}ner and A.~Rusetsky,
  \emph{{Energy shift of the three-particle system in a finite volume}},
  \href{https://doi.org/10.1103/PhysRevD.99.074513}{\emph{Phys. Rev. D}
  {\bfseries 99} (2019) 074513}
  [\href{https://arxiv.org/abs/1902.01111}{{\ttfamily 1902.01111}}].

\bibitem{Jackura:2019bmu}
A.W.~Jackura, S.M.~Dawid, C.~Fern\'andez-Ram\'\i{}rez, V.~Mathieu,
  M.~Mikhasenko, A.~Pilloni et~al., \emph{{Equivalence of three-particle
  scattering formalisms}},
  \href{https://doi.org/10.1103/PhysRevD.100.034508}{\emph{Phys. Rev. D}
  {\bfseries 100} (2019) 034508}
  [\href{https://arxiv.org/abs/1905.12007}{{\ttfamily 1905.12007}}].

\bibitem{Blanton:2019igq}
T.D.~Blanton, F.~Romero-{L\'{o}pez} and S.R.~Sharpe, \emph{{Implementing the
  three-particle quantization condition including higher partial waves}},
  \href{https://arxiv.org/abs/1901.07095}{{\ttfamily 1901.07095}}.

\bibitem{Briceno:2019muc}
R.A.~Brice\~no, M.T.~Hansen, S.R.~Sharpe and A.P.~Szczepaniak, \emph{{Unitarity
  of the infinite-volume three-particle scattering amplitude arising from a
  finite-volume formalism}},
  \href{https://doi.org/10.1103/PhysRevD.100.054508}{\emph{Phys. Rev. D}
  {\bfseries 100} (2019) 054508}
  [\href{https://arxiv.org/abs/1905.11188}{{\ttfamily 1905.11188}}].

\bibitem{Romero-Lopez:2019qrt}
F.~Romero-L\'opez, S.R.~Sharpe, T.D.~Blanton, R.A.~Brice\~no and M.T.~Hansen,
  \emph{{Numerical exploration of three relativistic particles in a finite
  volume including two-particle resonances and bound states}},
  \href{https://doi.org/10.1007/JHEP10(2019)007}{\emph{JHEP} {\bfseries 10}
  (2019) 007} [\href{https://arxiv.org/abs/1908.02411}{{\ttfamily
  1908.02411}}].

\bibitem{Pang:2020pkl}
J.-Y.~Pang, J.-J.~Wu and L.-S.~Geng, \emph{{$DDK$ system in finite volume}},
  \href{https://doi.org/10.1103/PhysRevD.102.114515}{\emph{Phys. Rev. D}
  {\bfseries 102} (2020) 114515}
  [\href{https://arxiv.org/abs/2008.13014}{{\ttfamily 2008.13014}}].

\bibitem{Blanton:2020gha}
T.D.~Blanton and S.R.~Sharpe, \emph{{Alternative derivation of the relativistic
  three-particle quantization condition}},
  \href{https://doi.org/10.1103/PhysRevD.102.054520}{\emph{Phys. Rev. D}
  {\bfseries 102} (2020) 054520}
  [\href{https://arxiv.org/abs/2007.16188}{{\ttfamily 2007.16188}}].

\bibitem{Hansen:2020zhy}
M.T.~Hansen, F.~Romero-L\'opez and S.R.~Sharpe, \emph{{Generalizing the
  relativistic quantization condition to include all three-pion isospin
  channels}}, \href{https://doi.org/10.1007/JHEP07(2020)047}{\emph{JHEP}
  {\bfseries 07} (2020) 047}
  [\href{https://arxiv.org/abs/2003.10974}{{\ttfamily 2003.10974}}].

\bibitem{Blanton:2020jnm}
T.D.~Blanton and S.R.~Sharpe, \emph{{Equivalence of relativistic three-particle
  quantization conditions}},
  \href{https://doi.org/10.1103/PhysRevD.102.054515}{\emph{Phys. Rev. D}
  {\bfseries 102} (2020) 054515}
  [\href{https://arxiv.org/abs/2007.16190}{{\ttfamily 2007.16190}}].

\bibitem{Romero-Lopez:2020rdq}
F.~Romero-L\'opez, A.~Rusetsky, N.~Schlage and C.~Urbach, \emph{{Relativistic
  $N$-particle energy shift in finite volume}},
  \href{https://doi.org/10.1007/JHEP02(2021)060}{\emph{JHEP} {\bfseries 02}
  (2021) 060} [\href{https://arxiv.org/abs/2010.11715}{{\ttfamily
  2010.11715}}].

\bibitem{Hansen:2021ofl}
M.T.~Hansen, F.~Romero-L\'opez and S.R.~Sharpe, \emph{{Decay amplitudes to
  three hadrons from finite-volume matrix elements}},
  \href{https://doi.org/10.1007/JHEP04(2021)113}{\emph{JHEP} {\bfseries 04}
  (2021) 113} [\href{https://arxiv.org/abs/2101.10246}{{\ttfamily
  2101.10246}}].

\bibitem{Blanton:2020gmf}
T.D.~Blanton and S.R.~Sharpe, \emph{{Relativistic three-particle quantization
  condition for nondegenerate scalars}},
  \href{https://doi.org/10.1103/PhysRevD.103.054503}{\emph{Phys. Rev. D}
  {\bfseries 103} (2021) 054503}
  [\href{https://arxiv.org/abs/2011.05520}{{\ttfamily 2011.05520}}].

\bibitem{Muller:2020vtt}
F.~M\"uller, T.~Yu and A.~Rusetsky, \emph{{Finite-volume energy shift of the
  three-pion ground state}},
  \href{https://doi.org/10.1103/PhysRevD.103.054506}{\emph{Phys. Rev. D}
  {\bfseries 103} (2021) 054506}
  [\href{https://arxiv.org/abs/2011.14178}{{\ttfamily 2011.14178}}].

\bibitem{Blanton:2021mih}
T.D.~Blanton and S.R.~Sharpe, \emph{{Three-particle finite-volume formalism for
  \ensuremath{\pi}+\ensuremath{\pi}+K+ and related systems}},
  \href{https://doi.org/10.1103/PhysRevD.104.034509}{\emph{Phys. Rev. D}
  {\bfseries 104} (2021) 034509}
  [\href{https://arxiv.org/abs/2105.12094}{{\ttfamily 2105.12094}}].

\bibitem{Muller:2021uur}
F.~M\"uller, J.-Y.~Pang, A.~Rusetsky and J.-J.~Wu,
  \emph{{Relativistic-invariant formulation of the NREFT three-particle
  quantization condition}},
  \href{https://doi.org/10.1007/JHEP02(2022)158}{\emph{JHEP} {\bfseries 02}
  (2022) 158} [\href{https://arxiv.org/abs/2110.09351}{{\ttfamily
  2110.09351}}].

\bibitem{Muller:2022oyw}
F.~M\"uller, J.-Y.~Pang, A.~Rusetsky and J.-J.~Wu, \emph{{Three-particle
  Lellouch-L\"uscher formalism in moving frames}},
  \href{https://doi.org/10.1007/JHEP02(2023)214}{\emph{JHEP} {\bfseries 02}
  (2023) 214} [\href{https://arxiv.org/abs/2211.10126}{{\ttfamily
  2211.10126}}].

\bibitem{Pang:2023jri}
J.-Y.~Pang, R.~Bubna, F.~M\"uller, A.~Rusetsky and J.-J.~Wu,
  \emph{{Lellouch-L\"uscher factor for the K \textrightarrow{}
  3\ensuremath{\pi} decays}},
  \href{https://doi.org/10.1007/JHEP05(2024)269}{\emph{JHEP} {\bfseries 05}
  (2024) 269} [\href{https://arxiv.org/abs/2312.04391}{{\ttfamily
  2312.04391}}].

\bibitem{Bubna:2023oxo}
R.~Bubna, F.~M\"uller and A.~Rusetsky, \emph{{Finite-volume energy shift of the
  three-nucleon ground state}},
  \href{https://doi.org/10.1103/PhysRevD.108.014518}{\emph{Phys. Rev. D}
  {\bfseries 108} (2023) 014518}
  [\href{https://arxiv.org/abs/2304.13635}{{\ttfamily 2304.13635}}].

\bibitem{Briceno:2024txg}
R.A.~Brice\~no, A.W.~Jackura, D.A.~Pefkou and F.~Romero-L\'opez,
  \emph{{Electroweak three-body decays in the presence of two- and three-body
  bound states}}, \href{https://doi.org/10.1007/JHEP05(2024)279}{\emph{JHEP}
  {\bfseries 05} (2024) 279}
  [\href{https://arxiv.org/abs/2402.12167}{{\ttfamily 2402.12167}}].

\bibitem{Xiao:2024dyw}
Q.-C.~Xiao, J.-Y.~Pang and J.-J.~Wu, \emph{{Lattice spectra of $DDK$ three-body
  system with Lorentz covariant kinematic}},
  \href{https://arxiv.org/abs/2408.16590}{{\ttfamily 2408.16590}}.

\bibitem{Hansen:2024ffk}
M.T.~Hansen, F.~Romero-L\'opez and S.R.~Sharpe, \emph{{Incorporating
  DD\ensuremath{\pi} effects and left-hand cuts in lattice QCD studies of the
  T$_{cc}$(3875)$^{+}$}},
  \href{https://doi.org/10.1007/JHEP06(2024)051}{\emph{JHEP} {\bfseries 06}
  (2024) 051} [\href{https://arxiv.org/abs/2401.06609}{{\ttfamily
  2401.06609}}].

\bibitem{Draper:2023xvu}
Z.T.~Draper, M.T.~Hansen, F.~Romero-L\'opez and S.R.~Sharpe, \emph{{Three
  relativistic neutrons in a finite volume}},
  \href{https://arxiv.org/abs/2303.10219}{{\ttfamily 2303.10219}}.

\bibitem{Feng:2024wyg}
Y.~Feng, F.~Gil, M.~D\"oring, R.~Molina, M.~Mai, V.~Shastry et~al., \emph{{A
  unitary coupled-channel three-body amplitude with pions and kaons}},
  \href{https://arxiv.org/abs/2407.08721}{{\ttfamily 2407.08721}}.

\bibitem{Jackura:2023qtp}
A.W.~Jackura and R.A.~Brice\~no, \emph{{Partial-wave projection of the
  one-particle exchange in three-body scattering amplitudes}},
  \href{https://doi.org/10.1103/PhysRevD.109.096030}{\emph{Phys. Rev. D}
  {\bfseries 109} (2024) 096030}
  [\href{https://arxiv.org/abs/2312.00625}{{\ttfamily 2312.00625}}].

\bibitem{Muller:2020wjo}
F.~M\"uller and A.~Rusetsky, \emph{{On the three-particle analog of the
  Lellouch-L\"uscher formula}},
  \href{https://doi.org/10.1007/JHEP03(2021)152}{\emph{JHEP} {\bfseries 03}
  (2021) 152} [\href{https://arxiv.org/abs/2012.13957}{{\ttfamily
  2012.13957}}].

\bibitem{Draper:2024qeh}
Z.T.~Draper and S.R.~Sharpe, \emph{{Three-particle formalism for multiple
  channels: the \ensuremath{\eta}\ensuremath{\pi}\ensuremath{\pi} + $
  K\overline{K}\pi $ system in isosymmetric QCD}},
  \href{https://doi.org/10.1007/JHEP07(2024)083}{\emph{JHEP} {\bfseries 07}
  (2024) 083} [\href{https://arxiv.org/abs/2403.20064}{{\ttfamily
  2403.20064}}].

\bibitem{Briceno:2024ehy}
R.A.~Brice\~no, C.S.R.~Costa and A.W.~Jackura, \emph{{Partial-wave projection
  of relativistic three-body amplitudes}},
  \href{https://doi.org/10.1103/PhysRevD.111.036029}{\emph{Phys. Rev. D}
  {\bfseries 111} (2025) 036029}
  [\href{https://arxiv.org/abs/2409.15577}{{\ttfamily 2409.15577}}].

\bibitem{Hansen:2019nir}
M.T.~Hansen and S.R.~Sharpe, \emph{{Lattice QCD and Three-particle Decays of
  Resonances}},  \href{https://arxiv.org/abs/1901.00483}{{\ttfamily
  1901.00483}}.

\bibitem{Rusetsky:2019gyk}
A.~Rusetsky, \emph{{Three particles on the lattice}},
  \href{https://doi.org/10.22323/1.363.0281}{\emph{PoS} {\bfseries LATTICE2019}
  (2019) 281} [\href{https://arxiv.org/abs/1911.01253}{{\ttfamily
  1911.01253}}].

\bibitem{Mai:2021lwb}
M.~Mai, M.~D\"oring and A.~Rusetsky, \emph{{Multi-particle systems on the
  lattice and chiral extrapolations: a brief review}},
  \href{https://doi.org/10.1140/epjs/s11734-021-00146-5}{\emph{Eur. Phys. J.
  ST} {\bfseries 230} (2021) 1623}
  [\href{https://arxiv.org/abs/2103.00577}{{\ttfamily 2103.00577}}].

\bibitem{Romero-Lopez:2022usb}
F.~Romero-L\'opez, \emph{{Multi-hadron interactions from lattice QCD}},
  \href{https://doi.org/10.22323/1.430.0235}{\emph{PoS} {\bfseries LATTICE2022}
  (2023) 235} [\href{https://arxiv.org/abs/2212.13793}{{\ttfamily
  2212.13793}}].

\bibitem{Beane:2007es}
S.R.~Beane, W.~Detmold, T.C.~Luu, K.~Orginos, M.J.~Savage and A.~Torok,
  \emph{{Multi-Pion Systems in Lattice QCD and the Three-Pion Interaction}},
  \href{https://doi.org/10.1103/PhysRevLett.100.082004}{\emph{Phys. Rev. Lett.}
  {\bfseries 100} (2008) 082004}
  [\href{https://arxiv.org/abs/0710.1827}{{\ttfamily 0710.1827}}].

\bibitem{Detmold:2008fn}
W.~Detmold, M.J.~Savage, A.~Torok, S.R.~Beane, T.C.~Luu, K.~Orginos et~al.,
  \emph{{Multi-Pion States in Lattice QCD and the Charged-Pion Condensate}},
  \href{https://doi.org/10.1103/PhysRevD.78.014507}{\emph{Phys. Rev. D}
  {\bfseries 78} (2008) 014507}
  [\href{https://arxiv.org/abs/0803.2728}{{\ttfamily 0803.2728}}].

\bibitem{Detmold:2008yn}
W.~Detmold, K.~Orginos, M.J.~Savage and A.~Walker-Loud, \emph{{Kaon
  Condensation with Lattice QCD}},
  \href{https://doi.org/10.1103/PhysRevD.78.054514}{\emph{Phys. Rev. D}
  {\bfseries 78} (2008) 054514}
  [\href{https://arxiv.org/abs/0807.1856}{{\ttfamily 0807.1856}}].

\bibitem{Detmold:2011kw}
W.~Detmold and B.~Smigielski, \emph{{Lattice QCD study of mixed systems of
  pions and kaons}},
  \href{https://doi.org/10.1103/PhysRevD.84.014508}{\emph{Phys. Rev. D}
  {\bfseries 84} (2011) 014508}
  [\href{https://arxiv.org/abs/1103.4362}{{\ttfamily 1103.4362}}].

\bibitem{Mai:2018djl}
M.~Mai and M.~Doring, \emph{{Finite-Volume Spectrum of $\pi^+\pi^+$ and
  $\pi^+\pi^+\pi^+$ Systems}},
  \href{https://doi.org/10.1103/PhysRevLett.122.062503}{\emph{Phys. Rev. Lett.}
  {\bfseries 122} (2019) 062503}
  [\href{https://arxiv.org/abs/1807.04746}{{\ttfamily 1807.04746}}].

\bibitem{Horz:2019rrn}
B.~H\"orz and A.~Hanlon, \emph{{Two- and three-pion finite-volume spectra at
  maximal isospin from lattice QCD}},
  \href{https://doi.org/10.1103/PhysRevLett.123.142002}{\emph{Phys. Rev. Lett.}
  {\bfseries 123} (2019) 142002}
  [\href{https://arxiv.org/abs/1905.04277}{{\ttfamily 1905.04277}}].

\bibitem{Blanton:2019vdk}
T.D.~Blanton, F.~Romero-L\'opez and S.R.~Sharpe, \emph{{$I=3$ Three-Pion
  Scattering Amplitude from Lattice QCD}},
  \href{https://doi.org/10.1103/PhysRevLett.124.032001}{\emph{Phys. Rev. Lett.}
  {\bfseries 124} (2020) 032001}
  [\href{https://arxiv.org/abs/1909.02973}{{\ttfamily 1909.02973}}].

\bibitem{Culver:2019vvu}
C.~Culver, M.~Mai, R.~Brett, A.~Alexandru and M.~D\"oring, \emph{{Three pion
  spectrum in the $I=3$ channel from lattice QCD}},
  \href{https://doi.org/10.1103/PhysRevD.101.114507}{\emph{Phys. Rev. D}
  {\bfseries 101} (2020) 114507}
  [\href{https://arxiv.org/abs/1911.09047}{{\ttfamily 1911.09047}}].

\bibitem{Mai:2019fba}
M.~Mai, M.~D\"oring, C.~Culver and A.~Alexandru, \emph{{Three-body unitarity
  versus finite-volume $\pi^+\pi^+\pi^+$ spectrum from lattice QCD}},
  \href{https://doi.org/10.1103/PhysRevD.101.054510}{\emph{Phys. Rev. D}
  {\bfseries 101} (2020) 054510}
  [\href{https://arxiv.org/abs/1909.05749}{{\ttfamily 1909.05749}}].

\bibitem{Fischer:2020jzp}
M.~Fischer, B.~Kostrzewa, L.~Liu, F.~Romero-L\'opez, M.~Ueding and C.~Urbach,
  \emph{{Scattering of two and three physical pions at maximal isospin from
  lattice QCD}},
  \href{https://doi.org/10.1140/epjc/s10052-021-09206-5}{\emph{Eur. Phys. J. C}
  {\bfseries 81} (2021) 436}
  [\href{https://arxiv.org/abs/2008.03035}{{\ttfamily 2008.03035}}].

\bibitem{Hansen:2020otl}
{\scshape Hadron Spectrum} collaboration, \emph{{Energy-Dependent $\pi^+ \pi^+
  \pi^+$ Scattering Amplitude from QCD}},
  \href{https://doi.org/10.1103/PhysRevLett.126.012001}{\emph{Phys. Rev. Lett.}
  {\bfseries 126} (2021) 012001}
  [\href{https://arxiv.org/abs/2009.04931}{{\ttfamily 2009.04931}}].

\bibitem{Alexandru:2020xqf}
A.~Alexandru, R.~Brett, C.~Culver, M.~D\"oring, D.~Guo, F.X.~Lee et~al.,
  \emph{{Finite-volume energy spectrum of the $K^-K^-K^-$ system}},
  \href{https://doi.org/10.1103/PhysRevD.102.114523}{\emph{Phys. Rev. D}
  {\bfseries 102} (2020) 114523}
  [\href{https://arxiv.org/abs/2009.12358}{{\ttfamily 2009.12358}}].

\bibitem{Brett:2021wyd}
R.~Brett, C.~Culver, M.~Mai, A.~Alexandru, M.~D\"oring and F.X.~Lee,
  \emph{{Three-body interactions from the finite-volume QCD spectrum}},
  \href{https://doi.org/10.1103/PhysRevD.104.014501}{\emph{Phys. Rev. D}
  {\bfseries 104} (2021) 014501}
  [\href{https://arxiv.org/abs/2101.06144}{{\ttfamily 2101.06144}}].

\bibitem{Blanton:2021llb}
T.D.~Blanton, A.D.~Hanlon, B.~H\"orz, C.~Morningstar, F.~Romero-L\'opez and
  S.R.~Sharpe, \emph{{Interactions of two and three mesons including higher
  partial waves from lattice QCD}},
  \href{https://doi.org/10.1007/JHEP10(2021)023}{\emph{JHEP} {\bfseries 10}
  (2021) 023} [\href{https://arxiv.org/abs/2106.05590}{{\ttfamily
  2106.05590}}].

\bibitem{NPLQCD:2020ozd}
{\scshape NPLQCD, QCDSF} collaboration, \emph{{Charged multihadron systems in
  lattice QCD+QED}},
  \href{https://doi.org/10.1103/PhysRevD.103.054504}{\emph{Phys. Rev. D}
  {\bfseries 103} (2021) 054504}
  [\href{https://arxiv.org/abs/2003.12130}{{\ttfamily 2003.12130}}].

\bibitem{Baeza-Ballesteros:2022bsn}
J.~Baeza-Ballesteros and M.T.~Hansen, \emph{{Two- and three-particle scattering
  in the (1+1)-dimensional O(3) non-linear sigma model}},
  \href{https://doi.org/10.22323/1.430.0050}{\emph{PoS} {\bfseries LATTICE2022}
  (2023) 050} [\href{https://arxiv.org/abs/2212.10623}{{\ttfamily
  2212.10623}}].

\bibitem{Draper:2023boj}
Z.T.~Draper, A.D.~Hanlon, B.~H\"orz, C.~Morningstar, F.~Romero-L\'opez and
  S.R.~Sharpe, \emph{{Interactions of \ensuremath{\pi}K,
  \ensuremath{\pi}\ensuremath{\pi}K and KK\ensuremath{\pi} systems at maximal
  isospin from lattice QCD}},
  \href{https://doi.org/10.1007/JHEP05(2023)137}{\emph{JHEP} {\bfseries 05}
  (2023) 137} [\href{https://arxiv.org/abs/2302.13587}{{\ttfamily
  2302.13587}}].

\bibitem{Abbott:2023coj}
{\scshape NPLQCD} collaboration, \emph{{Lattice quantum chromodynamics at large
  isospin density}},
  \href{https://doi.org/10.1103/PhysRevD.108.114506}{\emph{Phys. Rev. D}
  {\bfseries 108} (2023) 114506}
  [\href{https://arxiv.org/abs/2307.15014}{{\ttfamily 2307.15014}}].

\bibitem{Abbott:2024vhj}
{\scshape NPLQCD} collaboration, \emph{{QCD Constraints on Isospin-Dense Matter
  and the Nuclear Equation of State}},
  \href{https://doi.org/10.1103/PhysRevLett.134.011903}{\emph{Phys. Rev. Lett.}
  {\bfseries 134} (2025) 011903}
  [\href{https://arxiv.org/abs/2406.09273}{{\ttfamily 2406.09273}}].

\bibitem{Mai:2021nul}
{\scshape GWQCD} collaboration, \emph{{Three-Body Dynamics of the a1(1260)
  Resonance from Lattice QCD}},
  \href{https://doi.org/10.1103/PhysRevLett.127.222001}{\emph{Phys. Rev. Lett.}
  {\bfseries 127} (2021) 222001}
  [\href{https://arxiv.org/abs/2107.03973}{{\ttfamily 2107.03973}}].

\bibitem{Garofalo:2022pux}
M.~Garofalo, M.~Mai, F.~Romero-L\'opez, A.~Rusetsky and C.~Urbach,
  \emph{{Three-body resonances in the \ensuremath{\varphi}$^{4}$ theory}},
  \href{https://doi.org/10.1007/JHEP02(2023)252}{\emph{JHEP} {\bfseries 02}
  (2023) 252} [\href{https://arxiv.org/abs/2211.05605}{{\ttfamily
  2211.05605}}].

\bibitem{Yan:2024gwp}
H.~Yan, M.~Garofalo, M.~Mai, U.-G.~Mei\ss{}ner and C.~Urbach, \emph{{The
  $\omega$-meson from lattice QCD}},
  \href{https://arxiv.org/abs/2407.16659}{{\ttfamily 2407.16659}}.

\bibitem{Dawid:2025doq}
S.M.~Dawid, Z.T.~Draper, A.D.~Hanlon, B.~H\"orz, C.~Morningstar,
  F.~Romero-L\'opez et~al., \emph{{Two- and three-meson scattering amplitudes
  with physical quark masses from lattice QCD}},
  \href{https://arxiv.org/abs/2502.17976}{{\ttfamily 2502.17976}}.

\bibitem{Dawid:2025zxc}
S.M.~Dawid, Z.T.~Draper, A.D.~Hanlon, B.~H\"orz, C.~Morningstar,
  F.~Romero-L\'opez et~al., \emph{{QCD predictions for physical multimeson
  scattering amplitudes}},  \href{https://arxiv.org/abs/2502.14348}{{\ttfamily
  2502.14348}}.

\bibitem{Dawid:2024dgy}
S.M.~Dawid, F.~Romero-L\'opez and S.R.~Sharpe, \emph{{Finite- and
  infinite-volume study of $DD\pi$ scattering}},
  \href{https://arxiv.org/abs/2409.17059}{{\ttfamily 2409.17059}}.

\bibitem{Luscher:1986pf}
M.~Luscher, \emph{{Volume Dependence of the Energy Spectrum in Massive Quantum
  Field Theories. 2. Scattering States}},
  \href{https://doi.org/10.1007/BF01211097}{\emph{Commun. Math. Phys.}
  {\bfseries 105} (1986) 153}.

\bibitem{Luscher:1990ux}
M.~{L\"{u}scher}, \emph{{Two particle states on a torus and their relation to
  the scattering matrix}},
  \href{https://doi.org/10.1016/0550-3213(91)90366-6}{\emph{Nucl. Phys.}
  {\bfseries B354} (1991) 531}.

\bibitem{Green:2025rel}
J.R.~Green, \emph{{Status of two-baryon scattering in lattice QCD}},  in
  \emph{{11th International Workshop on Chiral Dynamics}}, 2, 2025
  [\href{https://arxiv.org/abs/2502.15546}{{\ttfamily 2502.15546}}].

\bibitem{MorningstarCD}
C.~Morningstar, \emph{{Nucleon resonances from lattice QCD}},  in \emph{{11th
  International Workshop on Chiral Dynamics}}, 2025.

\bibitem{Aoki:2025abn}
S.~Aoki, \emph{{Interaction between two hadrons in lattice QCD}},  in
  \emph{{11th International Workshop on Chiral Dynamics}}, 2, 2025
  [\href{https://arxiv.org/abs/2502.20671}{{\ttfamily 2502.20671}}].

\bibitem{Fischer:2020yvw}
{\scshape Extended Twisted Mass, ETM} collaboration, \emph{{The
  \ensuremath{\rho}-resonance from Nf=2 lattice QCD including the physical pion
  mass}}, \href{https://doi.org/10.1016/j.physletb.2021.136449}{\emph{Phys.
  Lett. B} {\bfseries 819} (2021) 136449}
  [\href{https://arxiv.org/abs/2006.13805}{{\ttfamily 2006.13805}}].

\bibitem{Alexandrou:2023elk}
C.~Alexandrou, S.~Bacchio, G.~Koutsou, T.~Leontiou, S.~Paul, M.~Petschlies
  et~al., \emph{{Elastic nucleon-pion scattering amplitudes in the
  \ensuremath{\Delta} channel at physical pion mass from lattice QCD}},
  \href{https://doi.org/10.1103/PhysRevD.109.034509}{\emph{Phys. Rev. D}
  {\bfseries 109} (2024) 034509}
  [\href{https://arxiv.org/abs/2307.12846}{{\ttfamily 2307.12846}}].

\bibitem{Bruno:2023pde}
M.~Bruno, D.~Hoying, T.~Izubuchi, C.~Lehner, A.S.~Meyer and M.~Tomii,
  \emph{{Isospin 0 and 2 two-pion scattering at physical pion mass using
  distillation with periodic boundary conditions in lattice QCD}},
  \href{https://arxiv.org/abs/2304.03313}{{\ttfamily 2304.03313}}.

\bibitem{RBC:2023xqv}
{\scshape RBC, UKQCD} collaboration, \emph{{Isospin 0 and 2 two-pion scattering
  at physical pion mass using all-to-all propagators with periodic boundary
  conditions in lattice QCD}},
  \href{https://doi.org/10.1103/PhysRevD.107.094512}{\emph{Phys. Rev. D}
  {\bfseries 107} (2023) 094512}
  [\href{https://arxiv.org/abs/2301.09286}{{\ttfamily 2301.09286}}].

\bibitem{Boyle:2024hvv}
P.~Boyle, F.~Erben, V.~G\"ulpers, M.T.~Hansen, F.~Joswig, M.~Marshall et~al.,
  \emph{{Light and Strange Vector Resonances from Lattice QCD at Physical Quark
  Masses}}, \href{https://doi.org/10.1103/PhysRevLett.134.111901}{\emph{Phys.
  Rev. Lett.} {\bfseries 134} (2025) 111901}
  [\href{https://arxiv.org/abs/2406.19194}{{\ttfamily 2406.19194}}].

\bibitem{Boyle:2024grr}
P.~Boyle, F.~Erben, V.~G\"ulpers, M.T.~Hansen, F.~Joswig, M.~Marshall et~al.,
  \emph{{Physical-mass calculation of \ensuremath{\rho}(770) and K*(892)
  resonance parameters via \ensuremath{\pi}\ensuremath{\pi} and
  K\ensuremath{\pi} scattering amplitudes from lattice QCD}},
  \href{https://doi.org/10.1103/PhysRevD.111.054510}{\emph{Phys. Rev. D}
  {\bfseries 111} (2025) 054510}
  [\href{https://arxiv.org/abs/2406.19193}{{\ttfamily 2406.19193}}].

\bibitem{Green:2021qol}
J.R.~Green, A.D.~Hanlon, P.M.~Junnarkar and H.~Wittig, \emph{{Weakly Bound H
  Dibaryon from SU(3)-Flavor-Symmetric QCD}},
  \href{https://doi.org/10.1103/PhysRevLett.127.242003}{\emph{Phys. Rev. Lett.}
  {\bfseries 127} (2021) 242003}
  [\href{https://arxiv.org/abs/2103.01054}{{\ttfamily 2103.01054}}].

\bibitem{Jackura:2020bsk}
A.W.~Jackura, R.A.~Brice\~no, S.M.~Dawid, M.H.E.~Islam and C.~McCarty,
  \emph{{Solving relativistic three-body integral equations in the presence of
  bound states}},
  \href{https://doi.org/10.1103/PhysRevD.104.014507}{\emph{Phys. Rev. D}
  {\bfseries 104} (2021) 014507}
  [\href{https://arxiv.org/abs/2010.09820}{{\ttfamily 2010.09820}}].

\bibitem{Dawid:2023jrj}
S.M.~Dawid, M.H.E.~Islam and R.A.~Brice\~no, \emph{{Analytic continuation of
  the relativistic three-particle scattering amplitudes}},
  \href{https://doi.org/10.1103/PhysRevD.108.034016}{\emph{Phys. Rev. D}
  {\bfseries 108} (2023) 034016}
  [\href{https://arxiv.org/abs/2303.04394}{{\ttfamily 2303.04394}}].

\bibitem{Dawid:2023kxu}
S.M.~Dawid, M.H.E.~Islam, R.A.~Briceno and A.W.~Jackura, \emph{{Evolution of
  Efimov states}},
  \href{https://doi.org/10.1103/PhysRevA.109.043325}{\emph{Phys. Rev. A}
  {\bfseries 109} (2024) 043325}
  [\href{https://arxiv.org/abs/2309.01732}{{\ttfamily 2309.01732}}].

\bibitem{Mohler:2017wnb}
D.~Mohler, S.~Schaefer and J.~Simeth, \emph{{CLS 2+1 flavor simulations at
  physical light- and strange-quark masses}},  in \emph{{35th International
  Symposium on Lattice Field Theory (Lattice 2017) Granada, Spain, June 18-24,
  2017}}, 2017,
  \href{https://inspirehep.net/record/1643031/files/arXiv:1712.04884.pdf}{https://inspirehep.net/record/1643031/files/arXiv:1712.04884.pdf}
  [\href{https://arxiv.org/abs/1712.04884}{{\ttfamily 1712.04884}}].

\bibitem{Luscher:1990ck}
M.~Luscher and U.~Wolff, \emph{{How to Calculate the Elastic Scattering Matrix
  in Two-dimensional Quantum Field Theories by Numerical Simulation}},
  \href{https://doi.org/10.1016/0550-3213(90)90540-T}{\emph{Nucl. Phys.}
  {\bfseries B339} (1990) 222}.

\bibitem{Hackett:2024xnx}
D.C.~Hackett and M.L.~Wagman, \emph{{Lanczos for lattice QCD matrix elements}},
   \href{https://arxiv.org/abs/2407.21777}{{\ttfamily 2407.21777}}.

\bibitem{Wagman:2024rid}
M.L.~Wagman, \emph{{Lanczos, the transfer matrix, and the signal-to-noise
  problem}},  \href{https://arxiv.org/abs/2406.20009}{{\ttfamily 2406.20009}}.

\bibitem{Ostmeyer:2024qgu}
J.~Ostmeyer, A.~Sen and C.~Urbach, \emph{{On the equivalence of Prony and
  Lanczos methods for Euclidean correlation functions}},
  \href{https://doi.org/10.1140/epja/s10050-025-01495-8}{\emph{Eur. Phys. J. A}
  {\bfseries 61} (2025) 26} [\href{https://arxiv.org/abs/2411.14981}{{\ttfamily
  2411.14981}}].

\bibitem{Hackett:2024nbe}
D.C.~Hackett and M.L.~Wagman, \emph{{Block Lanczos for lattice QCD spectroscopy
  and matrix elements}},  \href{https://arxiv.org/abs/2412.04444}{{\ttfamily
  2412.04444}}.

\bibitem{Abbott:2025yhm}
R.~Abbott, D.C.~Hackett, G.T.~Fleming, D.A.~Pefkou and M.L.~Wagman,
  \emph{{Filtered Rayleigh-Ritz is all you need}},
  \href{https://arxiv.org/abs/2503.17357}{{\ttfamily 2503.17357}}.

\bibitem{Garcia-Martin:2011iqs}
R.~Garcia-Martin, R.~Kaminski, J.R.~Pelaez, J.~Ruiz~de Elvira and
  F.J.~Yndurain, \emph{{The Pion-pion scattering amplitude. IV: Improved
  analysis with once subtracted Roy-like equations up to 1100 MeV}},
  \href{https://doi.org/10.1103/PhysRevD.83.074004}{\emph{Phys. Rev. D}
  {\bfseries 83} (2011) 074004}
  [\href{https://arxiv.org/abs/1102.2183}{{\ttfamily 1102.2183}}].

\bibitem{Pelaez:2020gnd}
J.R.~Pel\'aez and A.~Rodas, \emph{{Dispersive \ensuremath{\pi
  K}\textrightarrow{}\ensuremath{\pi K} and
  \ensuremath{\pi}\ensuremath{\pi}\textrightarrow{} \ensuremath{K\bar
  K}amplitudes from scattering data, threshold parameters, and the lightest
  strange resonance \ensuremath{\kappa} or \ensuremath{K_0^*}(700)}},
  \href{https://doi.org/10.1016/j.physrep.2022.03.004}{\emph{Phys. Rept.}
  {\bfseries 969} (2022) 1} [\href{https://arxiv.org/abs/2010.11222}{{\ttfamily
  2010.11222}}].

\bibitem{Bijnens:2021hpq}
J.~Bijnens and T.~Husek, \emph{{Six-pion amplitude}},
  \href{https://doi.org/10.1103/PhysRevD.104.054046}{\emph{Phys. Rev. D}
  {\bfseries 104} (2021) 054046}
  [\href{https://arxiv.org/abs/2107.06291}{{\ttfamily 2107.06291}}].

\bibitem{Bijnens:2022zsq}
J.~Bijnens, T.~Husek and M.~Sj\"o, \emph{{Six-meson amplitude in QCD-like
  theories}}, \href{https://doi.org/10.1103/PhysRevD.106.054021}{\emph{Phys.
  Rev. D} {\bfseries 106} (2022) 054021}
  [\href{https://arxiv.org/abs/2206.14212}{{\ttfamily 2206.14212}}].

\bibitem{Baeza-Ballesteros:2023ljl}
J.~Baeza-Ballesteros, J.~Bijnens, T.~Husek, F.~Romero-L\'opez, S.R.~Sharpe and
  M.~Sj\"o, \emph{{The isospin-3 three-particle K-matrix at NLO in ChPT}},
  \href{https://doi.org/10.1007/JHEP05(2023)187}{\emph{JHEP} {\bfseries 05}
  (2023) 187} [\href{https://arxiv.org/abs/2303.13206}{{\ttfamily
  2303.13206}}].

\bibitem{Baeza-Ballesteros:2024mii}
J.~Baeza-Ballesteros, J.~Bijnens, T.~Husek, F.~Romero-L\'opez, S.R.~Sharpe and
  M.~Sj\"o, \emph{{The three-pion K-matrix at NLO in ChPT}},
  \href{https://doi.org/10.1007/JHEP03(2024)048}{\emph{JHEP} {\bfseries 03}
  (2024) 048} [\href{https://arxiv.org/abs/2401.14293}{{\ttfamily
  2401.14293}}].

\bibitem{Padmanath:2022cvl}
M.~Padmanath and S.~Prelovsek, \emph{{Signature of a Doubly Charm Tetraquark
  Pole in DD* Scattering on the Lattice}},
  \href{https://doi.org/10.1103/PhysRevLett.129.032002}{\emph{Phys. Rev. Lett.}
  {\bfseries 129} (2022) 032002}
  [\href{https://arxiv.org/abs/2202.10110}{{\ttfamily 2202.10110}}].

\bibitem{Lyu:2023xro}
Y.~Lyu, S.~Aoki, T.~Doi, T.~Hatsuda, Y.~Ikeda and J.~Meng, \emph{{Doubly
  Charmed Tetraquark Tcc+ from Lattice QCD near Physical Point}},
  \href{https://doi.org/10.1103/PhysRevLett.131.161901}{\emph{Phys. Rev. Lett.}
  {\bfseries 131} (2023) 161901}
  [\href{https://arxiv.org/abs/2302.04505}{{\ttfamily 2302.04505}}].

\bibitem{Chen:2022vpo}
S.~Chen, C.~Shi, Y.~Chen, M.~Gong, Z.~Liu, W.~Sun et~al., \emph{{Tcc+(3875)
  relevant DD\textasteriskcentered{} scattering from Nf=2 lattice QCD}},
  \href{https://doi.org/10.1016/j.physletb.2022.137391}{\emph{Phys. Lett. B}
  {\bfseries 833} (2022) 137391}
  [\href{https://arxiv.org/abs/2206.06185}{{\ttfamily 2206.06185}}].

\bibitem{Ortiz-Pacheco:2023ble}
E.~Ortiz-Pacheco, S.~Collins, L.~Leskovec, M.~Padmanath and S.~Prelovsek,
  \emph{{Doubly charmed tetraquark: isospin channels and diquark-antidiquark
  interpolators}}, \href{https://doi.org/10.22323/1.453.0052}{\emph{PoS}
  {\bfseries LATTICE2023} (2024) 052}
  [\href{https://arxiv.org/abs/2312.13441}{{\ttfamily 2312.13441}}].

\bibitem{Collins:2024sfi}
S.~Collins, A.~Nefediev, M.~Padmanath and S.~Prelovsek, \emph{{Toward the quark
  mass dependence of Tcc+ from lattice QCD}},
  \href{https://doi.org/10.1103/PhysRevD.109.094509}{\emph{Phys. Rev. D}
  {\bfseries 109} (2024) 094509}
  [\href{https://arxiv.org/abs/2402.14715}{{\ttfamily 2402.14715}}].

\bibitem{Whyte:2024ihh}
{\scshape Hadron Spectrum} collaboration, \emph{{Near-threshold states in
  coupled DD*-D*D* scattering from lattice QCD}},
  \href{https://doi.org/10.1103/PhysRevD.111.034511}{\emph{Phys. Rev. D}
  {\bfseries 111} (2025) 034511}
  [\href{https://arxiv.org/abs/2405.15741}{{\ttfamily 2405.15741}}].

\bibitem{Stump:2024lqx}
A.~Stump and J.R.~Green, \emph{{Distillation and position-space sampling for
  local multiquark interpolators}},
  \href{https://doi.org/10.22323/1.466.0094}{\emph{PoS} {\bfseries LATTICE2024}
  (2025) 094} [\href{https://arxiv.org/abs/2412.09246}{{\ttfamily
  2412.09246}}].

\bibitem{Vujmilovic:2024snz}
I.~Vujmilovic, S.~Collins, L.~Leskovec, E.~Ortiz-Pacheco, P.~Madanagopalan and
  S.~Prelovsek, \emph{{$T^+_{cc}$ via the plane wave approach and including
  diquark-antidiquark operators}},
  \href{https://doi.org/10.22323/1.466.0112}{\emph{PoS} {\bfseries LATTICE2024}
  (2025) 112} [\href{https://arxiv.org/abs/2411.08646}{{\ttfamily
  2411.08646}}].

\bibitem{Meng:2024kkp}
L.~Meng, E.~Ortiz-Pacheco, V.~Baru, E.~Epelbaum, M.~Padmanath and S.~Prelovsek,
  \emph{{Doubly charm tetraquark channel with isospin 1 from lattice QCD}},
  \href{https://doi.org/10.1103/PhysRevD.111.034509}{\emph{Phys. Rev. D}
  {\bfseries 111} (2025) 034509}
  [\href{https://arxiv.org/abs/2411.06266}{{\ttfamily 2411.06266}}].

\bibitem{Albaladejo:2021vln}
M.~Albaladejo, \emph{{Tcc+ coupled channel analysis and predictions}},
  \href{https://doi.org/10.1016/j.physletb.2022.137052}{\emph{Phys. Lett. B}
  {\bfseries 829} (2022) 137052}
  [\href{https://arxiv.org/abs/2110.02944}{{\ttfamily 2110.02944}}].

\bibitem{Du:2021zzh}
M.-L.~Du, V.~Baru, X.-K.~Dong, A.~Filin, F.-K.~Guo, C.~Hanhart et~al.,
  \emph{{Coupled-channel approach to Tcc+ including three-body effects}},
  \href{https://doi.org/10.1103/PhysRevD.105.014024}{\emph{Phys. Rev. D}
  {\bfseries 105} (2022) 014024}
  [\href{https://arxiv.org/abs/2110.13765}{{\ttfamily 2110.13765}}].

\bibitem{Achasov:2022onn}
N.N.~Achasov and G.N.~Shestakov, \emph{{Triangle singularities in the
  Tcc+\textrightarrow{}D*+D0\textrightarrow{}\ensuremath{\pi}+D0D0 decay
  width}}, \href{https://doi.org/10.1103/PhysRevD.105.096038}{\emph{Phys. Rev.
  D} {\bfseries 105} (2022) 096038}
  [\href{https://arxiv.org/abs/2203.17100}{{\ttfamily 2203.17100}}].

\bibitem{Wang:2023iaz}
J.-Z.~Wang, Z.-Y.~Lin and S.-L.~Zhu, \emph{{Cut structures and an observable
  singularity in the three-body threshold dynamics: The Tcc+ case}},
  \href{https://doi.org/10.1103/PhysRevD.109.L071505}{\emph{Phys. Rev. D}
  {\bfseries 109} (2024) L071505}
  [\href{https://arxiv.org/abs/2309.09861}{{\ttfamily 2309.09861}}].

\bibitem{Zhang:2024dth}
X.~Zhang, \emph{{Relativistic three-body scattering and the D0D*+-D+D*0
  system}}, \href{https://doi.org/10.1103/PhysRevD.109.094010}{\emph{Phys. Rev.
  D} {\bfseries 109} (2024) 094010}
  [\href{https://arxiv.org/abs/2402.02151}{{\ttfamily 2402.02151}}].

\bibitem{Du:2025vkm}
M.-L.~Du, F.-K.~Guo and B.~Wu, \emph{{Effective range expansion with the
  left-hand cut and its application to the $T_{cc}(3875)$}},  in \emph{{11th
  International Workshop on Chiral Dynamics}}, 2, 2025
  [\href{https://arxiv.org/abs/2502.19774}{{\ttfamily 2502.19774}}].

\bibitem{Abolnikov:2024key}
M.~Abolnikov, V.~Baru, E.~Epelbaum, A.A.~Filin, C.~Hanhart and L.~Meng,
  \emph{{Internal structure of the Tcc(3875)+ from its light-quark mass
  dependence}},
  \href{https://doi.org/10.1016/j.physletb.2024.139188}{\emph{Phys. Lett. B}
  {\bfseries 860} (2025) 139188}
  [\href{https://arxiv.org/abs/2407.04649}{{\ttfamily 2407.04649}}].

\bibitem{Meng:2023bmz}
L.~Meng, V.~Baru, E.~Epelbaum, A.A.~Filin and A.M.~Gasparyan, \emph{{Solving
  the left-hand cut problem in lattice QCD: Tcc(3875)+ from finite volume
  energy levels}},
  \href{https://doi.org/10.1103/PhysRevD.109.L071506}{\emph{Phys. Rev. D}
  {\bfseries 109} (2024) L071506}
  [\href{https://arxiv.org/abs/2312.01930}{{\ttfamily 2312.01930}}].

\bibitem{Du:2023hlu}
M.-L.~Du, A.~Filin, V.~Baru, X.-K.~Dong, E.~Epelbaum, F.-K.~Guo et~al.,
  \emph{{Role of Left-Hand Cut Contributions on Pole Extractions from Lattice
  Data: Case Study for Tcc(3875)+}},
  \href{https://doi.org/10.1103/PhysRevLett.131.131903}{\emph{Phys. Rev. Lett.}
  {\bfseries 131} (2023) 131903}
  [\href{https://arxiv.org/abs/2303.09441}{{\ttfamily 2303.09441}}].

\bibitem{Bubna:2024izx}
R.~Bubna, H.-W.~Hammer, F.~M\"uller, J.-Y.~Pang, A.~Rusetsky and J.-J.~Wu,
  \emph{{L\"uscher equation with long-range forces}},
  \href{https://doi.org/10.1007/JHEP05(2024)168}{\emph{JHEP} {\bfseries 05}
  (2024) 168} [\href{https://arxiv.org/abs/2402.12985}{{\ttfamily
  2402.12985}}].

\bibitem{Raposo:2023oru}
A.B.a.~Raposo and M.T.~Hansen, \emph{{Finite-volume scattering on the left-hand
  cut}}, \href{https://doi.org/10.1007/JHEP08(2024)075}{\emph{JHEP} {\bfseries
  08} (2024) 075} [\href{https://arxiv.org/abs/2311.18793}{{\ttfamily
  2311.18793}}].

\bibitem{Raposo:2025dkb}
A.B.a.~Raposo, R.A.~Brice\~no, M.T.~Hansen and A.W.~Jackura, \emph{{Extracting
  scattering amplitudes for arbitrary two-particle systems with one-particle
  left-hand cuts via lattice QCD}},
  \href{https://arxiv.org/abs/2502.19375}{{\ttfamily 2502.19375}}.

\bibitem{HansenCD}
M.T.~Hansen, \emph{{Left-hand branch cuts in lattice QCD scattering
  calculations}},  in \emph{{11th International Workshop on Chiral Dynamics}},
  2025.

\bibitem{Yan:2024yuq}
H.~Yan, C.~Liu, L.~Liu, Y.~Meng and H.~Xing, \emph{{Pion mass dependence in
  D\ensuremath{\pi} scattering and the D0*(2300) resonance from lattice QCD}},
  \href{https://doi.org/10.1103/PhysRevD.111.014503}{\emph{Phys. Rev. D}
  {\bfseries 111} (2025) 014503}
  [\href{https://arxiv.org/abs/2404.13479}{{\ttfamily 2404.13479}}].

\bibitem{Belyaev:1994zk}
V.M.~Belyaev, V.M.~Braun, A.~Khodjamirian and R.~Ruckl, \emph{{D* D pi and B* B
  pi couplings in QCD}},
  \href{https://doi.org/10.1103/PhysRevD.51.6177}{\emph{Phys. Rev. D}
  {\bfseries 51} (1995) 6177}
  [\href{https://arxiv.org/abs/hep-ph/9410280}{{\ttfamily hep-ph/9410280}}].

\bibitem{BaBar:2013zgp}
{\scshape BaBar} collaboration, \emph{{Measurement of the $D^*(2010)^+$ natural
  line width and the $D^*(2010)^+ - D^0$ mass difference}},
  \href{https://doi.org/10.1103/PhysRevD.88.052003}{\emph{Phys. Rev. D}
  {\bfseries 88} (2013) 052003}
  [\href{https://arxiv.org/abs/1304.5009}{{\ttfamily 1304.5009}}].

\end{thebibliography}\endgroup

\end{document}